\documentclass[aps, prl, twocolumn, superscriptaddress, longbibliography]{revtex4-1}
\usepackage[colorlinks, linkcolor=blue, anchorcolor=blue, citecolor=blue]{hyperref}
\usepackage{amsmath}
\usepackage{graphicx}
\usepackage{braket}
\usepackage{color}
\usepackage{soul}

\begin{document}

\title{Global phase diagram of a spin-orbital Kondo impurity model and the suppression of Fermi-liquid scale}
\author{Y. Wang}     
\affiliation{Department of Condensed Matter Physics and Materials Science, Brookhaven National Laboratory, Upton, New York 11973, USA} 

\author{E. Walter}
\affiliation{Arnold Sommerfeld Center for Theoretical Physics, 
Center for NanoScience,\looseness=-1\,  and Munich 
Center for \\ Quantum Science and Technology,\looseness=-2\, Ludwig-Maximilians-Universit\"at M\"unchen, 80333 Munich, Germany}
\author{S.-S. B. Lee}
\affiliation{Arnold Sommerfeld Center for Theoretical Physics, 
Center for NanoScience,\looseness=-1\,  and Munich 
Center for \\ Quantum Science and Technology,\looseness=-2\, Ludwig-Maximilians-Universit\"at M\"unchen, 80333 Munich, Germany}
\author{K. M. Stadler}
\affiliation{Arnold Sommerfeld Center for Theoretical Physics, 
Center for NanoScience,\looseness=-1\,  and Munich 
Center for \\ Quantum Science and Technology,\looseness=-2\, Ludwig-Maximilians-Universit\"at M\"unchen, 80333 Munich, Germany}
\author{J. von Delft}
\affiliation{Arnold Sommerfeld Center for Theoretical Physics, 
Center for NanoScience,\looseness=-1\,  and Munich 
Center for \\ Quantum Science and Technology,\looseness=-2\, Ludwig-Maximilians-Universit\"at M\"unchen, 80333 Munich, Germany}
\author{A. Weichselbaum}
\affiliation{Department of Condensed Matter Physics and Materials Science, Brookhaven National Laboratory, Upton, New York 11973, USA} 
\affiliation{Arnold Sommerfeld Center for Theoretical Physics, 
Center for NanoScience,\looseness=-1\,  and Munich 
Center for \\ Quantum Science and Technology,\looseness=-2\, Ludwig-Maximilians-Universit\"at M\"unchen, 80333 Munich, Germany}
\author{G. Kotliar}
\affiliation{Department of Condensed Matter Physics and Materials Science, Brookhaven National Laboratory, Upton, New York 11973, USA} 
\affiliation{Department of Physics and Astronomy, Rutgers University, Piscataway, New Jersey 08856, USA}

\date{\today}

\begin{abstract}
Many correlated metallic materials are described by Landau Fermi-liquid theory at low energies, but for Hund metals the Fermi-liquid coherence
scale $T_{\text{FL}}$ is found to be surprisingly small. In this Letter, we study the simplest impurity model relevant for Hund
metals, the three-channel spin-orbital Kondo model, using the numerical renormalization group (NRG) method and compute its global phase diagram. In this framework, $T_{\text{FL}}$  becomes arbitrarily small close to two new quantum critical points (QCPs) which we identify by tuning the spin or spin-orbital Kondo couplings into the ferromagnetic regimes.
We find quantum phase transitions to a singular Fermi-liquid or a novel non-Fermi-liquid phase. 
The new non-Fermi-liquid phase shows frustrated behavior involving alternating overscreenings in spin and orbital sectors, with universal power laws in the spin ($\omega^{-1/5}$), orbital ($\omega^{1/5}$) and spin-orbital ($\omega^1$) dynamical susceptibilities. These power laws, and the NRG eigenlevel spectra, can be fully understood using conformal field theory arguments, which also clarify the nature of the non-Fermi-liquid phase.
\end{abstract}

\maketitle
\textit{Introduction.}---A very large number of correlated metallic materials
exhibit broad regimes of temperature $T$ characterized by large values of resistivity exceeding the Mott-Ioffe-Regel limit~\cite{emery:1995}. At lower temperature, the magnitude of the resistivity decreases, but it does not follow the canonical Fermi-liquid (FL) $T^2$ behavior~\cite{landau:1980}. The Landau FL quasiparticles emerge only at very low temperature, below a coherence scale $T_{\text{FL}}$ which
is much smaller than the natural energy scales of the problem, set by the electronic hopppings. Hund metals~\cite{yin:2011,werner:2008,medici:2011,georges:2013,stadler:2019}, including ruthenates~\cite{tyler:1998,mackenzie:1998,schneider:2014,mravlje:2011,takeshi:2016,deng:2016}, iron pnictides and chalcogenides~\cite{Kamihara:2008,hardy:2013,yiming:2013,walmsley:2013,haule:2009,yin:2012}, are such class of materials. Why is $T_{\text{FL}}$ so small in units of the bandwidth? This ``naturalness problem" is a central problem of condensed matter physics which has attracted considerable attention in the community. Its solution should also provide a clue as to what reference system should be used to describe the anomalous behavior observed in a broad energy regime above $T_{\text{FL}}$, when no other instabilities such as superconductivity intervene.

A solution to this riddle has been proposed in terms of proximity to quantum critical points (QCPs)~\cite{Sachdev:2000,sachdev:2007}, signaling the transition to an ordered phase, or to an unconventional one such as a Mott insulator. An alternative starting point has been provided by the development of the dynamical mean field theory (DMFT)~\cite{georges:1996}. Here, the excitations of a solid are understood in terms of atomic multiplets embedded in an effective medium, and the evolution of the electronic structure from atomic multiplet excitations into quasiparticles arises naturally. This approach has been turned into an \textit{ab-initio} method (LDA+DMFT)~\cite{lichtenstein:2001,aichhorn:2009,haule:2010} and has provided quantitative predictions in many materials of interest~\cite{kotliar:2006,Shim:2007,haule:2008,haule:2009,Hauleurs:2009,aichhorn:2009,yin:2011,deng:2014,shinaoka:2015,kim:2018}, where \textit{ab-initio} calculations are in remarkable agreement with experiments. However, the solution of the LDA+DMFT equations 
is a complex problem, 
which generically yields a non-zero FL scale. Hence no connection with the ideas of QCPs was made.
The question of how to reduce the FL scale to exactly zero and how to characterize the ensuing anomalous behavior above $T_{\text{FL}}$ has remained open.  

In this Letter, we provide an answer to this question by computing a global phase diagram of the simplest three-channel spin-orbital Kondo model which captures the essential physics of Hund metals, using the exact numerical renormalization group (NRG) method~\cite{bulla:2008}. By tuning the spin or spin-orbital Kondo couplings into the ferromagnetic regimes, we push $T_{\text{FL}}$ to be exactly zero and identify QCPs. We find quantum phase transitions to a singular-Fermi-liquid (SFL) phase and to a novel non-Fermi-liquid (NFL) phase showing frustrated behavior of alternating overscreenings in spin and orbital sectors, with universal power laws in dynamical susceptibilities. We use conformal field theory (CFT) arguments~\cite{affleck:1990,affleck:1991a,affleck:1991b,affleck:1993,ludwig:1994,walter:2019} to identify the nature of the NFL phase, analytically reproduce the NRG eigenlevel spectra and explain the power laws. Our global phase diagram provides a clear picture for understanding the suppression of coherence in Hund metals in terms of proximity to QCPs.

\begin{figure}
\centering
\includegraphics[width=0.48\textwidth]{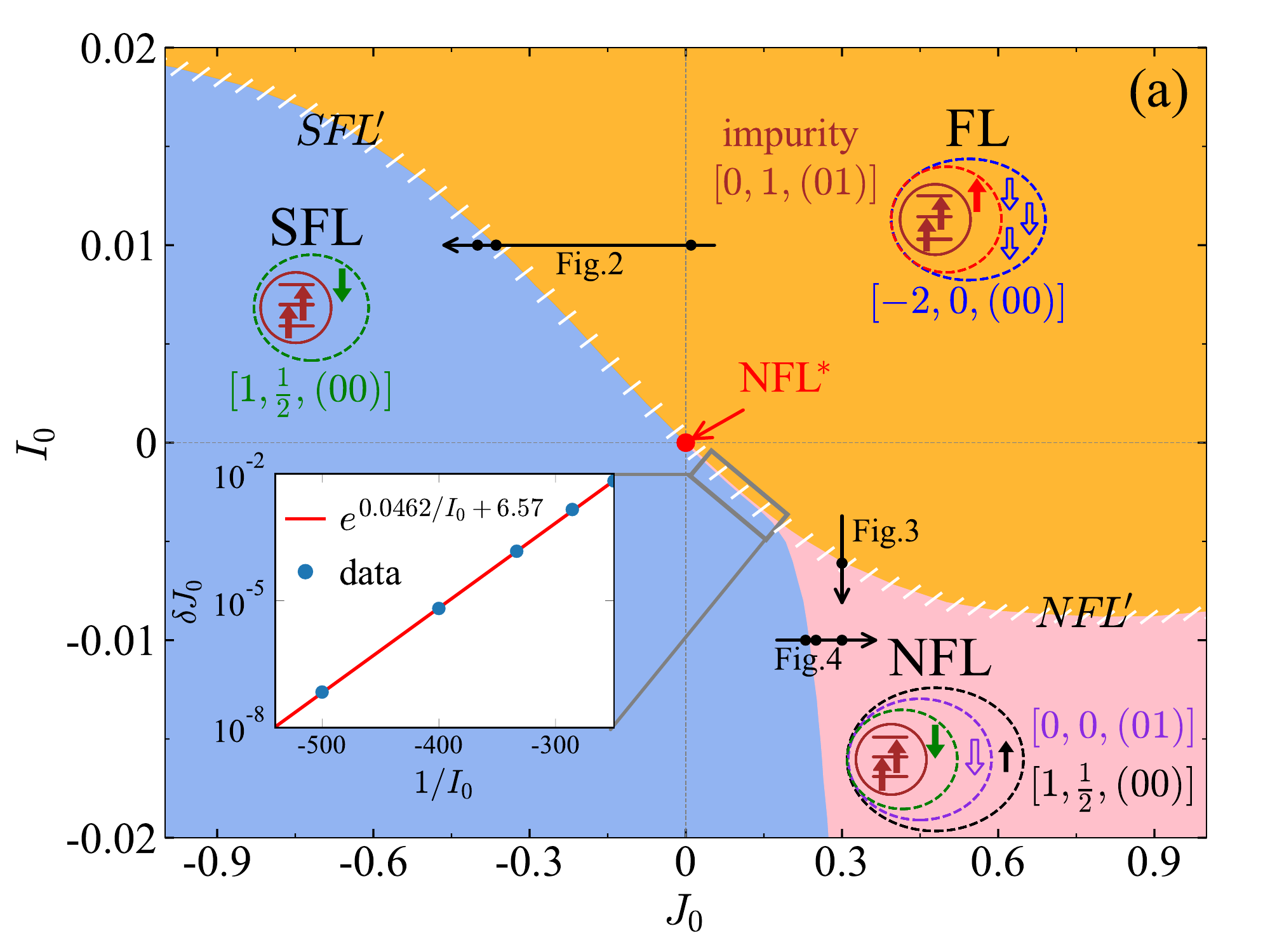}
\includegraphics[width=0.48\textwidth]{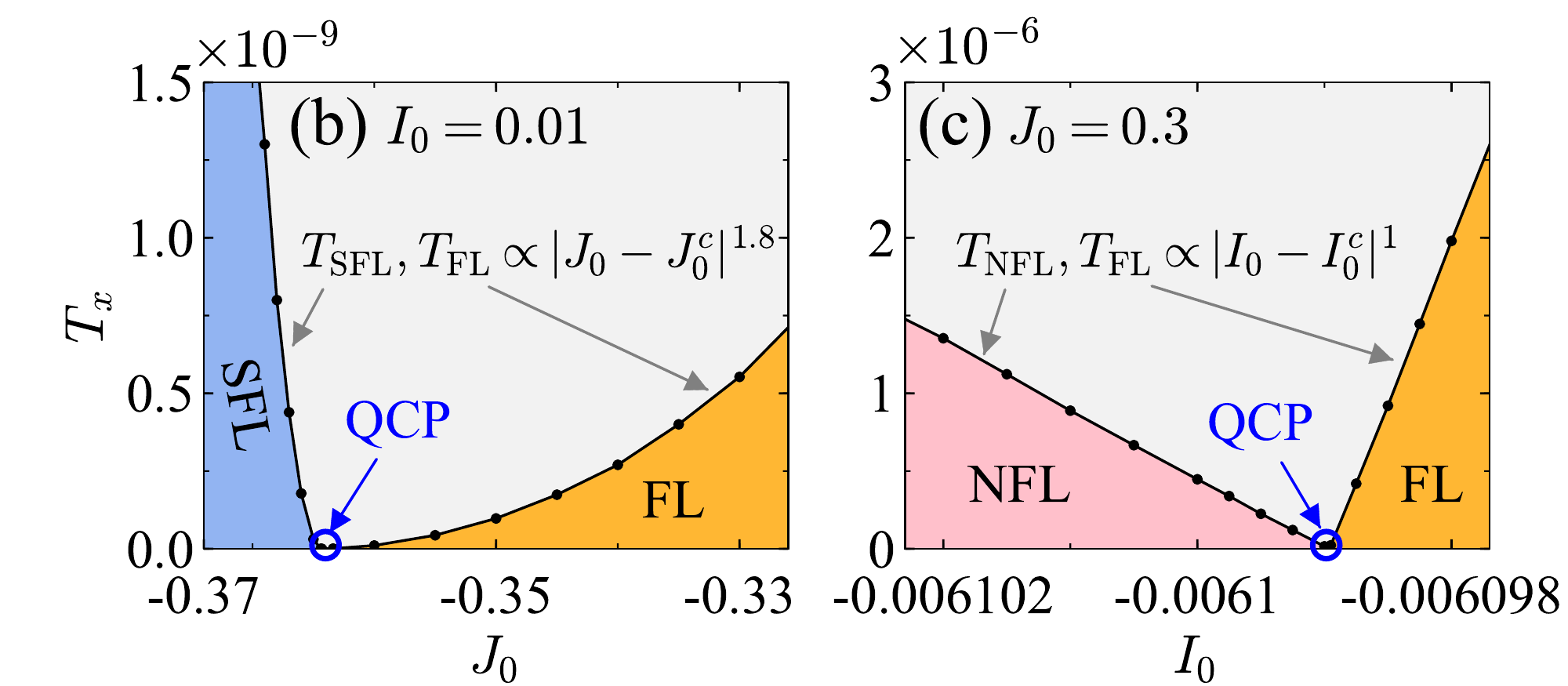}
\includegraphics[width=0.48\textwidth]{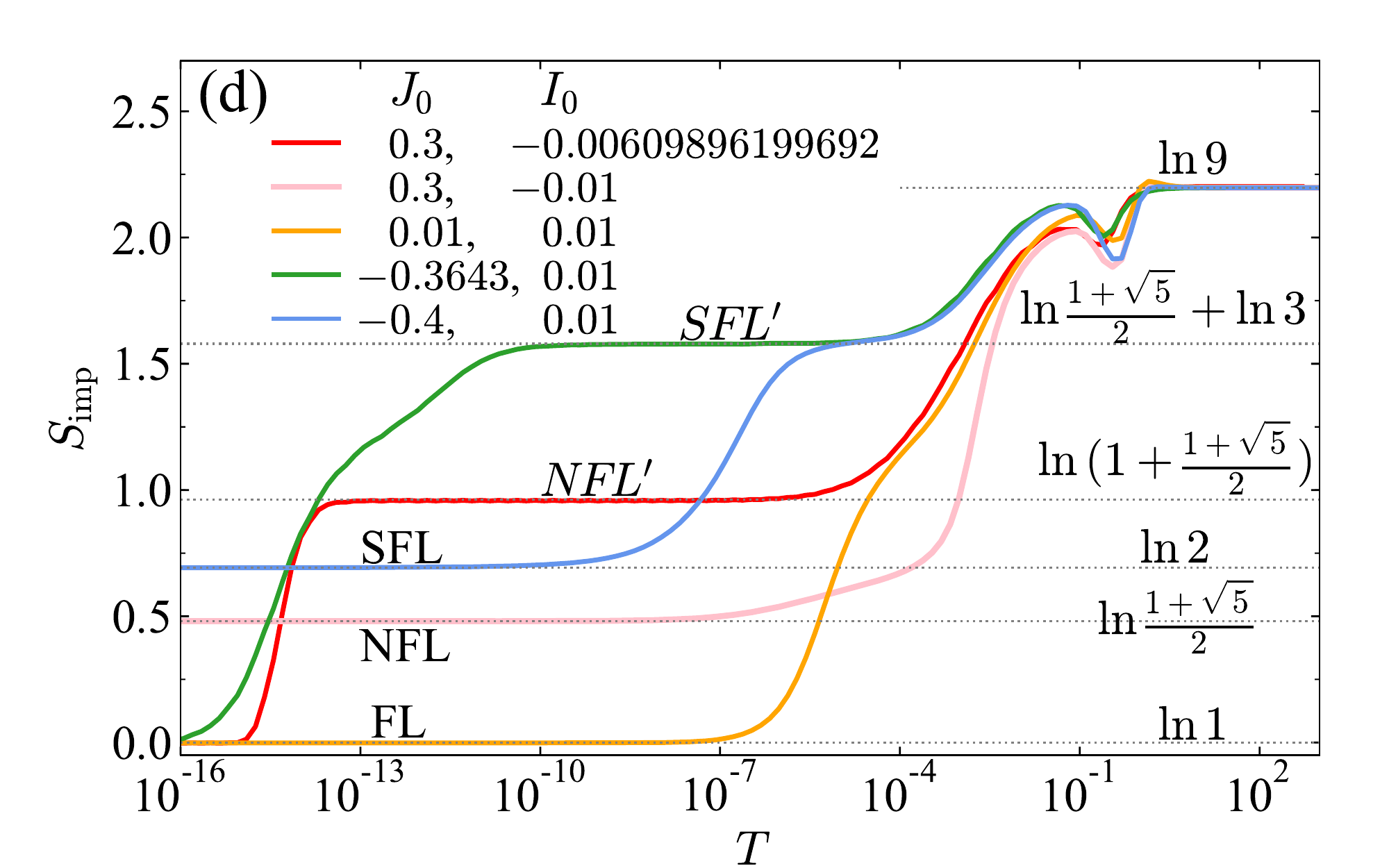}
\caption{(a) The calculated global phase diagram vs. $J_0$ and $I_{0}$ at fixed $K_0=0.3$. 
         Four low-energy fixed points are found:
         Fermi-liquid (FL, orange region);
         singular Fermi-liquid (SFL, blue region) 
         with underscreened spin and fully screened orbital isospin;
         frustrated non-Fermi-liquid (NFL, pink region) 
         with alternating spin and orbital overscreenings;
         and non-Fermi-liquid NFL$^{\!\ast}$ (red dot at $J_0=0, I_0=0$) 
         with overscreened orbital isospin and degenerate impurity spin $\tfrac12$, $\tfrac32$.
         Cartoons depict the respective screening processes,
         where one dashed ellipse loosely represents an even number of Wilson shells.
         The indicated additional charge then is relative to  half-filling, where
         filled (empty) arrows represent electrons (holes) with corresponding spin direction.
         The white-hatched region indicates the existence of an intermediate-energy crossover regime $\it{SFL}^{\!\prime}$ ($\it{NFL}^{\!\prime}$) enclosing the phase boundary between FL and SFL (NFL).
         The inset shows the ``funnel width", $\delta J_0$,
         of the NFL phase vs. $1/I_0$ when $I_0\rightarrow 0^{-}$. 
         (b,c) The onset energy scales $T_{x}$ for ($x=$) FL, SFL and NFL vs.\ (b) $J_0$ or (c) $I_0$, where quantum critical points are identified.
         (d) Impurity contribution to entropy $S_{\text{imp}}$ as functions of temperature $T$.}
\label{fig:phase}
\end{figure}

\textit{Model and Methods.}---We study the three-channel spin-orbital Kondo (3soK) model proposed in~\cite{yin:2012,aron:2015} for the  studies of Hund metals.
$H_{\text{bath}} = \sum_{pm\sigma}\varepsilon_{p} \psi^{\dagger}_{pm\sigma}\psi_{pm\sigma}$ describes a symmetric, flat-band bath with half-bandwidth $D=1$, where $\psi^{\dagger}_{pm\sigma}$ creates an electron with momentum $p$ and spin $\sigma$ in orbital $m\in \{1,2,3\}$. 
The bath couples to the impurity spin $\mathbf{S}$ and orbital isospin $\mathbf{T}$ via
\begin{equation}
\label{Eqn:int}
H_{\text{int}} = J_0 \mathbf{S}\cdot\mathbf{J}_{\text{sp}} + K_0\mathbf{T}\cdot\mathbf{J}_{\text{orb}} + I_0\mathbf{S}\cdot\mathbf{J}_{\text{sp-orb}}\cdot\mathbf{T}.
\end{equation}
Here $\mathbf{S}$ are SU(2) generators in the $S=1$ representation, normalized as $\text{Tr}(S^{\alpha}S^{\beta})=\frac{1}{2}\delta_{\alpha\beta}$,   and $\mathbf{T}$ are SU(3) generators in the
$\bar{3}$, i.e. $(01)$ representation~\cite{andreas:2012}, and $\text{Tr}(T^aT^b)=\frac{1}{2}\delta_{ab}$. $\mathbf{J}_{\text{sp}}$, $\mathbf{J}_{\text{orb}}$ and $\mathbf{J}_{\text{sp-orb}}$ are the bath spin, orbital and spin-orbital densities at the impurity site, with $J^{\alpha}_{\text{sp}}=\psi^{\dagger}_{m\sigma}\frac{1}{2}\sigma^{\alpha}_{\sigma\sigma^{\prime}}\psi_{m\sigma^{\prime}}$, $J^{a}_{\text{orb}}=\psi^{\dagger}_{m\sigma}\frac{1}{2}\tau^{a}_{m m^{\prime}}\psi_{m^{\prime}\sigma}$,
$J^{\alpha,a}_{\text{sp-orb}}=\frac{1}{4}  \psi^{\dagger}_{m\sigma}
\sigma^{\alpha}_{\sigma\sigma^{\prime}}\tau^{a}_{m m^{\prime}}\psi_{m^{\prime}\sigma^{\prime}}$ (summation over repeated indices is implied)
and normalized 
$\psi^{\dagger}_{m\sigma}{=}\tfrac{1}{\sqrt{N}}
\sum_{p}\psi^{\dagger}_{pm\sigma}$, and
$\sigma^{\alpha} [\tau^{a}]$ are Pauli [Gell-Mann] matrices, with normalization $\text{Tr}(\sigma^{\alpha}\sigma^{\beta})=2\delta_{\alpha\beta}$ [$\text{Tr}(\tau^a\tau^b)=2\delta_{ab}$].
$J_0$, $K_0$ and $I_0$ are bare spin, orbital and spin-orbital Kondo exchange couplings, with positive and negative values describing antiferromagnetic (AFM) and ferromagnetic (FM) couplings, respectively.
We take $K_0{=}0.3$ throughout.

We use the full-density-matrix NRG~\cite{andreas:2007} method to solve this model, exploiting its full
U(1)$_{\text{ch}}$ $\times$ SU(2)$_{\text{sp}}$ $\times$ SU(3)$_{\text{orb}}$ symmetry using QSpace~\cite{andreas:2012}. 
Symmetry labels $Q \equiv [q, S, (\lambda_1\lambda_2)]$ are used to label multiplets,
where $q$ is the bath particle number relative
to half-filling of the bath (we choose $q_{\text{imp}}=0$ because the impurity site has no charge dynamics), $S$ is the total spin,
and $(\lambda_1\lambda_2)$ labels a SU(3) representation described by a Young diagram with $\lambda_1+\lambda_2$ ($\lambda_2$) boxes in its first (second) row. The impurity multiplet has $Q_\mathrm{imp}{=}[0, 1, (01)]$. The bath is discretized logarithmically and mapped to a semi-infinite ``Wilson chain" with exponentially decaying hoppings, and the impurity coupled to chain site $k=0$. The chain is diagonalized iteratively while discarding high-energy states, thereby zooming in on low-energy properties:
the finite-size level spacing of a chain ending at site $k\ge 0$ 
is of order $\omega_k\propto \Lambda^{-k/2}$. Here 
$\Lambda>1$ is a discretization parameter, 
chosen to be 4 in this work. The RG flow can be visualized by combining the rescaled low-lying NRG eigenlevel spectra,
$E=(\mathcal{E}-\mathcal{E}_{\text{ref}})/\omega_{k}$ vs. $\omega_{k}$,
with increasing even or odd $k$.
The imaginary part of the impurity dynamical susceptibilities $\chi^{\text{imp}}_{\text{sp}}$,  $\chi^{\text{imp}}_{\text{orb}}$ and $\chi^{\text{imp}}_{\text{sp-orb}}$ were calculated at temperature $T=10^{-16}$. Computational details are presented in the Supplemental Material~\cite{suppl}.

\begin{figure}
\includegraphics[width=0.48\textwidth]{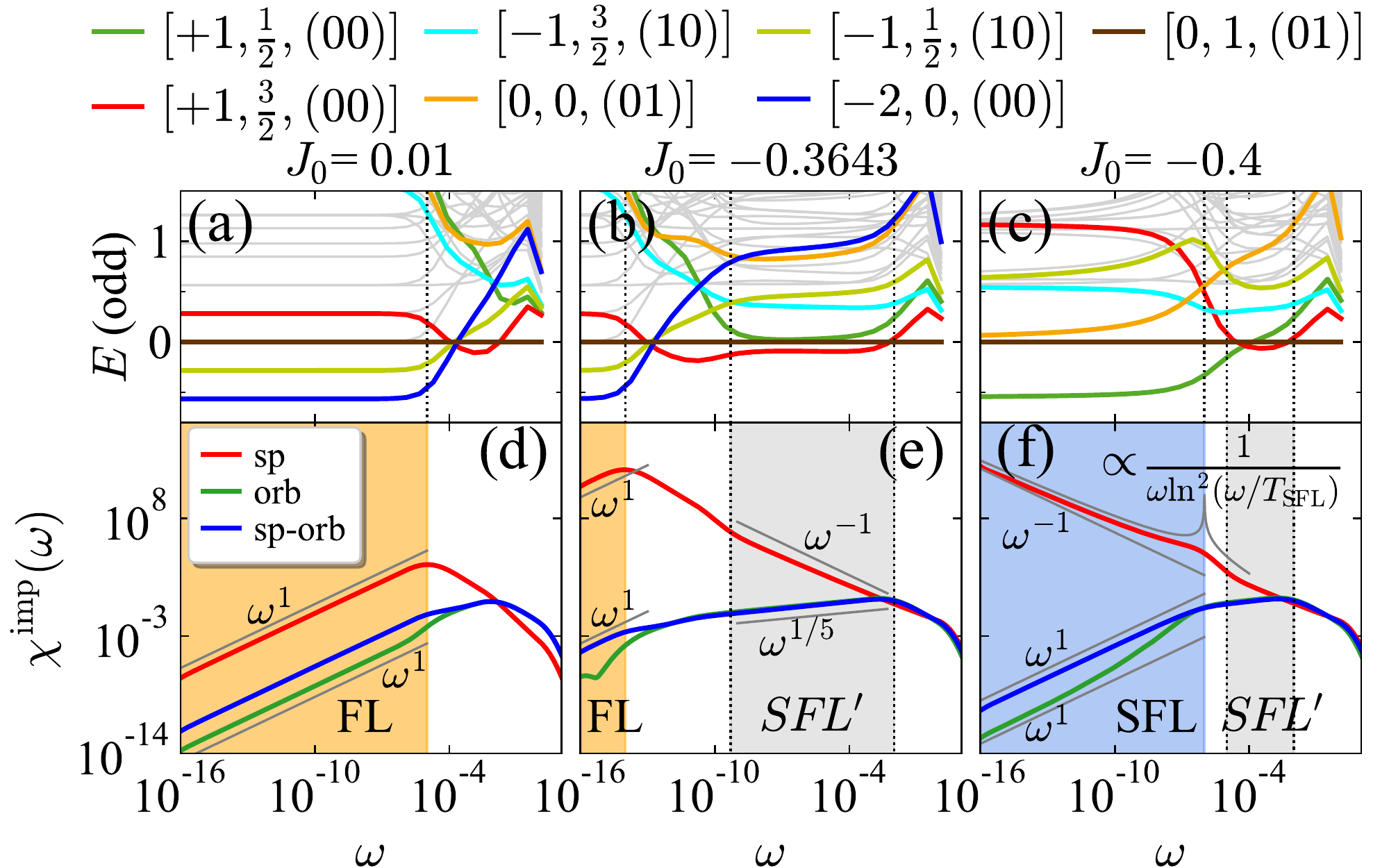}
\caption{The phase transition from FL to SFL at $I_0=0.01$.
         (a)-(c) NRG flow diagrams of a Wilson chain with odd 
         length $k$, with the energy of the lowest $[0, 1, (01)]$ multiplet as the reference energy $\mathcal{E}_{\text{ref}}$. The symmetry labels of selected multiplets are shown on top.
         (d)-(f) Impurity dynamical susceptibility $\chi^{\text{imp}}(\omega)$. 
         }
\label{fig:flow1}
\end{figure}

\textit{Fixed points.}---The calculated global phase diagram as a function of $J_0$ and $I_0$ is shown in Fig.~\ref{fig:phase}(a). We first describe the low-energy fixed points found in the phase diagram. Throughout the entire regions where all three Kondo couplings are AFM, and for part of regions where $J_0$ or $I_0$ takes FM values (orange region), the system flows to a low-energy FL fixed point. This is seen in the
NRG flow diagram and dynamical impurity susceptibilities $\chi^{\text{imp}}$ at $J_{0}=I_0=0.01$ in Figs.~\ref{fig:flow1}(a,d).
The ground state is a spin and orbital singlet, with impurity entropy $S_{\text{imp}}=\ln 1$ [orange curve in Fig.~\ref{fig:phase}(d)].
For small $\omega$, all $\chi^{\text{imp}}$ follow a $\omega$-linear behavior, characteristic of a FL. 

When $J_0$ takes FM values and $I_0$ FM or small AFM values (blue region), the phase is governed by a low-energy  SFL~\cite{coleman:2003,mehta:2005,koller:2005} fixed point  where the spin is underscreened while the orbitals are fully screened. 
The transition from FL to SFL is analyzed in Fig.~\ref{fig:flow1} for $I_0=0.01$.
%
Figs.~\ref{fig:flow1}(c,f), computed for $J_0=-0.4$, 
show the NRG flow and $\chi^{\text{imp}}$ to the SFL 
fixed point. It has ground state $[+1,\frac{1}{2},(00)]$ and $S_{\text{imp}}$ approaches
$\ln 2$ at low energies [blue curve in Fig.~\ref{fig:phase}(d)], signaling a residual spin of $\frac{1}{2}$. $\chi^{\text{imp}}_{\text{sp}}$ deviates slightly from a pure $\omega^{-1}$ power-law by a logarithmic correction at high energy and can be fitted by $\sim1/(\omega\ln^{2}(\omega/T_{\text{SFL}}))$ with $T_{\text{SFL}}$ as an onset energy scale, consistent with the SFL results in~\cite{koller:2005}. $\chi^{\text{imp}}_{\text{orb}}$ shows $\omega$-linear behavior at low energy, indicating fully screened orbital isospin. The coefficient of the impurity specific heat, $C_{\text{imp}}(T)/T$~\cite{suppl}, shows divergent behavior~\cite{coleman:2003}, confirming the singular nature of this fixed point.

\begin{figure}
\includegraphics[width=0.48\textwidth]{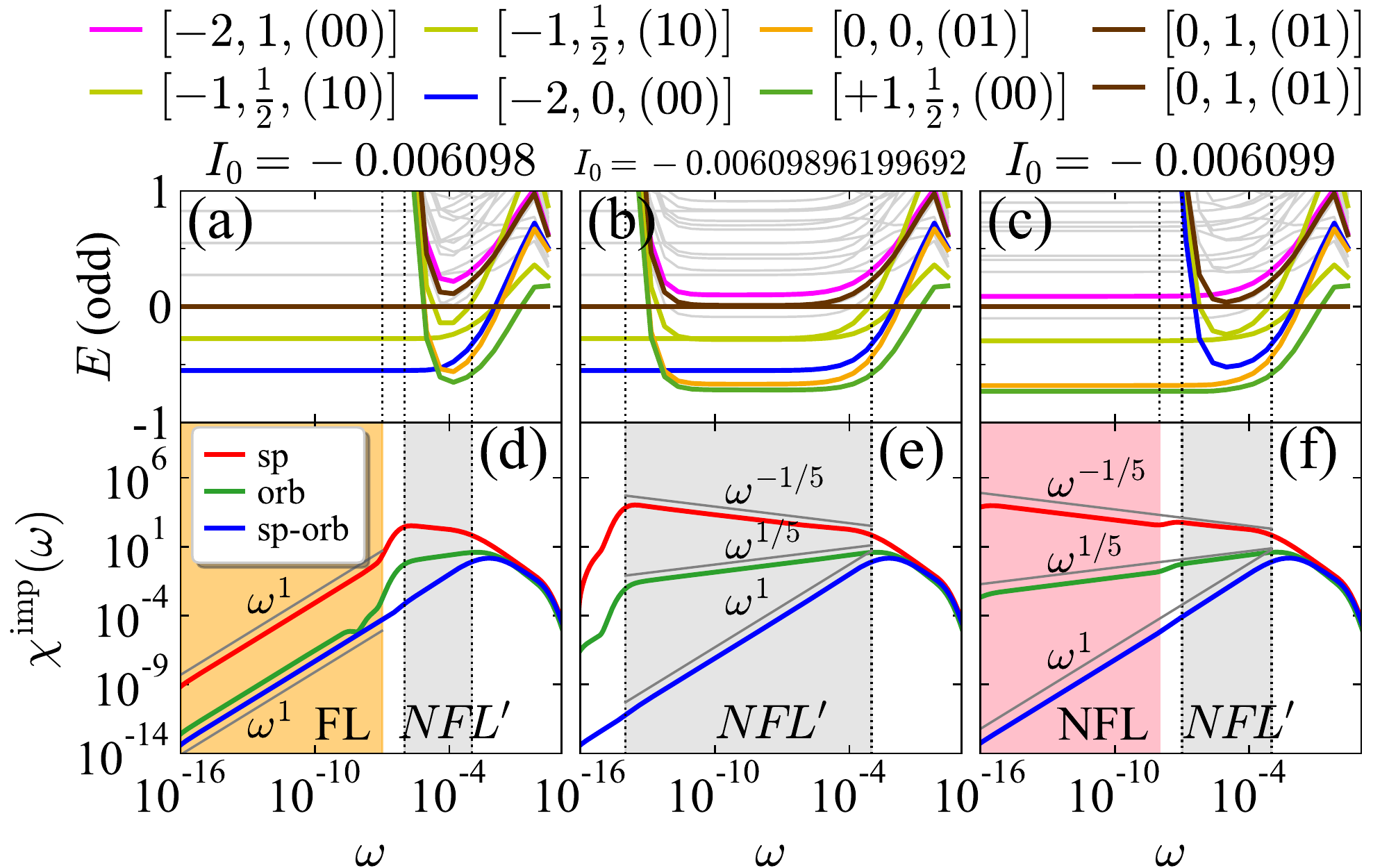}
\caption{Analogous to Fig.~\ref{fig:flow1}, but for the phase transition from FL to NFL at $J_0=0.3$.
}
\label{fig:flow2}
\end{figure}

\begin{figure}
\includegraphics[width=0.48\textwidth]{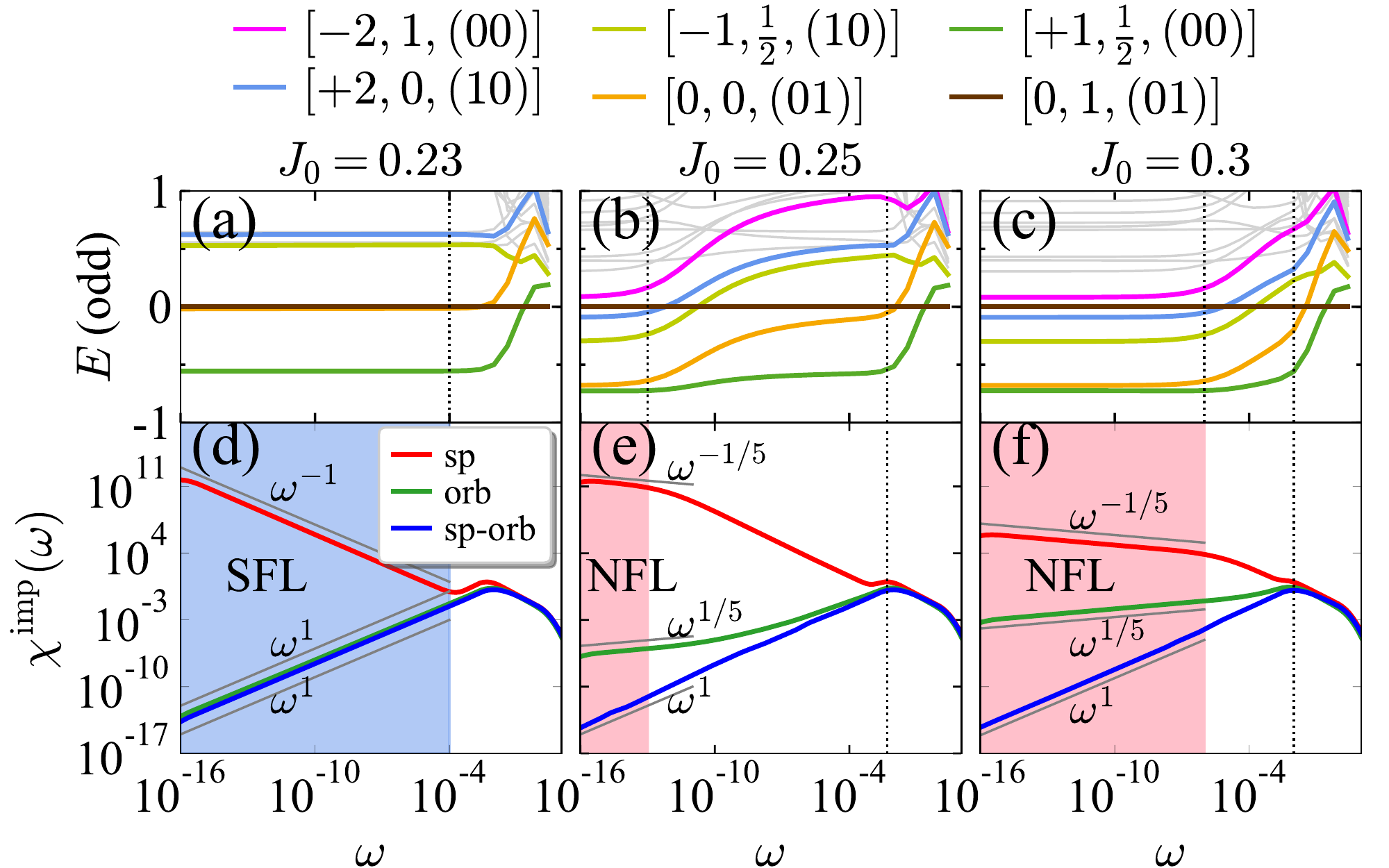}
\caption{Analogous to Fig.~\ref{fig:flow1}, but for the phase transition from SFL to NFL at $I_0=-0.01$.
}
\label{fig:flow3}
\end{figure}

When $I_0$ takes strong FM and $J_0$ strong AFM couplings (pink region), we find a novel NFL fixed point, showing very interesting frustrated behavior of alternating overscreenings in 
spin and orbital sectors.
\mbox{Figure~\ref{fig:flow2}} analyzes the transition from FL to NFL at $J_0=0.3$.
Figs.~\ref{fig:flow2}(c,f) and~\ref{fig:flow3}(c,f)
show the NRG flow 
and $\chi^{\text{imp}}$ towards the NFL fixed point. The two lowest multiplets with either orbital singlet, $[+1, \frac{1}{2}, (00)]$, or spin singlet, $[0, 0, (01)]$, are very close in energy. The dynamical susceptibilities follow perfect and universal power laws for the spin ($\omega^{-1/5}$), oribtal ($\omega^{1/5}$) and spin-orbital ($\omega^{1}$) operators. The impurity entropy $S_{\text{imp}}$ evaluates to $\ln \frac{1+\sqrt{5}}{2}$ [pink curve in Fig.~\ref{fig:phase}(d)]. This value can be obtained from Eq.(6) in~\cite{parcollet:1998} for a general SU($N$)$_K$ Kondo model ($K$ is the number of channels) with $N=3, K=2, Q=2$ indicating SU(3)$_2$ orbital overscreening, or 
with $N=2, K=3, Q=1$ indicating SU(2)$_3$ spin overscreening. Motivated by this, we follow the recently developed SU(2)$\times$SU(3) CFT approach~\cite{walter:2019} to identify the nature of this fixed point. Its NRG eigenlevel spectra $Q^{\prime}{=}[q^{\prime}, S^{\prime}, (\lambda_1^{\prime}\lambda_2^{\prime})]$ can be reproduced by applying either a SU(2)$_3$ fusion procedure in the spin sector or a SU(3)$_2$ fusion procedure in the orbital sector, i.e. fusing a spectrum of free fermion $Q=[q, S, (\lambda_1\lambda_2)]$, with an effective impurity multiplet labeling either $Q_{\text{imp}}^{\text{eff}}{=}[+1, \frac{1}{2}, (00)]$, or $Q_{\text{imp}}^{\text{eff}}{=}[0, 0, (01)]$. Double fusion of the spectrum $Q^{\prime}$ with the conjugate representation of the impurity multiplet, $\bar{Q}_{\text{imp}}^{\text{eff}}{=}[-1, \frac{1}{2}, (00)]$ or $\bar{Q}_{\text{imp}}^{\text{eff}}{=}[0, 0, (10)]$, yields the quantum numbers $Q^{\prime\prime}{=}[q^{\prime\prime}, S^{\prime\prime}, (\lambda_1^{\prime\prime}\lambda_2^{\prime\prime})]$ to characterize the CFT boundary operators, with scaling dimensions $\Delta$, determining the behavior of dynamical susceptibilities. 

Tables S1-S2 in the Supplemental Material~\cite{suppl} show the CFT results of the fixed point spectra and compare them with the NRG spectra at $J_0=0.3, I_0=-0.01$. Both fusion procedures yield the same results, which reproduce the NRG spectra very well. The scaling dimension of the leading boundary operator in the spin, orbital and spin-orbital sectors are found to be $\Delta_{\text{sp}}=\frac{2}{5}$, $\Delta_{\text{orb}}=\frac{3}{5}$ and $\Delta_{\text{sp-orb}}=1$, respectively. They are also consistent with the CFT results in~\cite{parcollet:1998} for either a spin SU(2)$_3$ Kondo model ($\Delta_{\text{sp}}=\frac{2}{2+3}$, $\Delta_{\text{orb}}=\frac{3}{2+3}$), or an orbital SU(3)$_2$  Kondo model ($\Delta_{\text{sp}}=\frac{2}{3+2}$, $\Delta_{\text{orb}}=\frac{3}{3+2}$). The power laws of dynamical susceptibilities can then be understood by the CFT procedure~\cite{walter:2019} $\chi^{\text{imp}}_{\text{sp}}\sim\omega^{2\Delta_{\text{sp}}-1}=\omega^{-1/5}$, $\chi^{\text{imp}}_{\text{orb}}\sim\omega^{2\Delta_{\text{orb}}-1}=\omega^{1/5}$ and $\chi^{\text{imp}}_{\text{sp-orb}}\sim\omega^{2\Delta_{\text{sp-orb}}-1}=\omega^{1}$, respectively. 

The impurity entropy and the CFT analysis both suggest that the spin SU(2)$_3$ and orbital SU(3)$_2$ Kondo models with overscreened fixed points are actually equivalent and complementary descriptions of this NFL fixed point. It indicates an alternating spin SU(2)$_3$ and orbital SU(3)$_2$ overscreening process by successively binding one electron or one hole, as illustrated by the cartoon picture at the bottom right of Fig.~\ref{fig:phase}(a), similar in spirit to that of Nozi\`eres and Blandin~\cite{nozieres:1980}.
To be specific, the strong AFM orbital coupling binds the bare impurity $Q_{\text{imp}}=[0,1,(01)]$ and one bath electron $[+1,\frac12,(10)]$ into a fully screened orbital singlet with either spin $\frac32$ or $\frac12$:
$[0,1,(01)] \otimes [+1, \tfrac{1}{2}, (10)] \rightarrow [+1,\tfrac{3}{2}, (00)] \oplus [+1,\tfrac{1}{2}, (00)] \,$.
In the FL phase, the spin $\frac32$ multiplet has the lower energy; it can then bind three holes to form a fully screened spin and orbital singlet \cite{walter:2019}:
$[+1,\tfrac32,(00)] \otimes [-3, \tfrac{3}{2}, (00)] \rightarrow [-2,0, (00)] \,$.
By contrast, in the NFL regime, the spin $\tfrac12$ multiplet has the lower energy since the spin-orbital coupling $I_0$ is strongly FM. Next, the AFM spin coupling attempts to screen the spin $\tfrac12$ by coupling it to one hole, to yield a spin singlet,
\begin{subequations}
\begin{equation}
[+1,\tfrac12,(00)] \otimes [-1,\tfrac12,(01)] \rightarrow [0,0,(01)] \,,
\end{equation}
but the result is an overscreened orbital isospin. Screening the latter by binding an electron,
\begin{equation}
[0,0,(01)] \otimes [+1,\tfrac12,(10)] \rightarrow [+1,\tfrac12,(00)] \,,
\end{equation}
\end{subequations}
leads back to an overscreened spin.
Overall, this results in a neverending alternation of spin and orbital overscreening, favored by the fact that the multiplets $[0, 0, (01)]$ and $[+1, \frac{1}{2}, (00)]$ are lowest in energy [see Figs.~\ref{fig:flow2}(c),~\ref{fig:flow3}(c)], with a very small energy difference.

The special point at $J_0=I_0=0$ 
corresponds to a SU(3)$_2$ NFL fixed point (NFL$^{\ast}$)
with overscreened orbitals and a degenerate impurity spin of $\tfrac12$, $\tfrac32$.
The inset of Fig.~\ref{fig:phase}(a) suggests that the region of NFL actually extends to this point. 
There we analyze the width of the NFL ``funnel", defined by $\delta J_0=J_{0}^{c1}-J_{0}^{c2}$, vs. $1/I_0$, where $J_{0}^{c1}$ ($J_{0}^{c2}$) is the phase boundary between FL (SFL) and NFL. It follows $\exp (0.0462/I_0+6.57)$, becoming zero only when $I_0\rightarrow 0^{-}$.

\textit{Phase transitions.}---$T_{\text{FL}}$ on the FL side and $T_{\text{SFL}}$ ($T_{\text{NFL}}$, the NFL scale) on the SFL (NFL) side go to zero as the phase boundary is approached.
We find that $T_{\text{FL}}$, $T_{\text{SFL}}$ and $T_{\text{NFL}}$ follow power laws as functions of the control parameters $J_0$ and $I_0$,  $|J_0-J_0^{c}|^{\alpha}$ and $|I_0-I_0^{c}|^{\alpha}$, to approach exactly zero at the critical values $J_0^{c}$ and $I_0^{c}$, signalling the existence of QCPs~\cite{Sachdev:2000,sachdev:2007}.
The exponents found are $\alpha{=}1.8$ in the FL-SFL transition, and $\alpha{=}1$ for FL-NFL. 
We show $T_{\text{FL/SFL}}$ as functions of $J_0$ at $I_0=0.01$ in Fig.~\ref{fig:phase}(b), and $T_{\text{FL/NFL}}$ as functions of $I_0$ at $J_0=0.3$ in Fig.~\ref{fig:phase}(c). More data are shown in Fig. S5~\cite{suppl}.

When approaching the QCP in the FL-SFL transition as in Fig.~\ref{fig:flow1} by decreasing $J_0$, 
the spin-orbital separation window~\cite{stadler:2015,stadler:2019}
increases a lot, 
as seen in Figs.~\ref{fig:flow1}(b,e) for $J_0=-0.3643$,
and a wide crossover regime, $\it{SFL}^{\prime}$, forms at intermediate energies. 
There the impurity entropy $S_{\text{imp}}$
evaluates to $\ln \frac{1+\sqrt{5}}{2}$ + $\ln 3$ [green curve in Fig.~\ref{fig:phase}(d)], corresponding to an orbital overscreened SU(3)$_2$ fixed point, coupled to a fluctuating spin-1 moment. 
This is consistent with the recent findings in the region $I_0=0$ and $J_0\rightarrow 0^{+}$ in~\cite{horvat:2019}. $\chi^{\text{imp}}_{\text{orb}}$ follows a universal power-law of $\omega^{1/5}$, showing similarity with the NFL phase due to the same orbital SU(3)$_2$ overscreening, while $\chi^{\text{imp}}_{\text{sp}}$ follows an approximate power law (with some non-power-law corrections, see~\cite{suppl}). 
Across the phase transition, the multiplet $[+1, \frac{1}{2}, (00)]$ is pushed down to be the new ground state, while the original ground state $[-2, 0, (00)]$ of the FL phase is pushed up to very high energy. 

When approaching the QCP in the FL-NFL transition as
illustrated in Fig.~\ref{fig:flow2}, fine-tuning of $I_0$ generates a large crossover regime $\it{NFL}^{\prime}$ at intermediate energies [Figs.~\ref{fig:flow2}(b,e)], where the set of low-lying states is simply the union of those of the FL and NFL spectra (see Table S4 in~\cite{suppl}). $\it{NFL}^{\prime}$ thus represents a ``level-crossing" scenario~\cite{sbierski:2013,lee:2019,suppl}, involving two orthogonal low-energy subspaces whose levels cross when $I_0$ is tuned. When sufficiently close,
both subspaces contribute to thermodynamic and dynamical properties. Here, the FL and NFL compete in the intermediate-energy regime, and $I_0$ determines either FL [Fig.~\ref{fig:flow2}(a,d)] or NFL [Fig.~\ref{fig:flow2}(c,f)] to be the low-energy fixed point. The impurity entropy $S_{\text{imp}}^{\it{NFL}^{\prime}}$ evaluates to $\ln (e^{S_{\text{imp}}^{\text{FL}}}+e^{S_{\text{imp}}^{\text{NFL}}})=\ln (1+\frac{1+\sqrt{5}}{2})$ [red curve in Fig.~\ref{fig:phase}(d)], not $\ln 1+\ln \frac{1+\sqrt{5}}{2}$,
because the FL and NFL subspaces do not overlap.
Hence the total effective impurity degrees of freedom is
the \textit{sum} of the contributions of those two sectors~\cite{suppl}. 
$\chi^{\text{imp}}$ of $\it{NFL}^{\prime}$ follow the same power laws as NFL because the NFL part dominates in this regime.
For more details on $\it{NFL}^{\prime}$, see~\cite{suppl}.

The transition from SFL to NFL shown in Fig.~\ref{fig:flow3} confirms the picture of alternating overscreenings. Tuning $J_0$ to be more AFM, the state $[0,0,(01)]$ is pushed down to be nearly degenerate with the ground state $[+1, \frac{1}{2},(00)]$ [Fig.~\ref{fig:flow3}(b)], signalling the start of the alternating overscreening process. $\chi^{\text{imp}}_{\text{sp}}$ bends downward away from the $\omega^{-1}$ behavior towards a $\omega^{-1/5}$ dependence, while $\chi^{\text{imp}}_{\text{orb}}$ bends upward away from the $\omega$-linear behavior towards a $\omega^{1/5}$ dependence. $\chi^{\text{imp}}_{\text{sp-orb}}$ still follows $\omega^1$.

\textit{Conclusion.}---To summarize, we have presented a global phase diagram of the 3soK model. This allows us to follow the suppression of the coherence scale in Hund metals down to zero energy. The new NFL phase contains the essential ingredients needed to understand the actual incoherent behavior seen above $T_{\text{FL}}$. 
Recent advances in the physics of cold atoms might actually offer a concrete realization of the phase diagram of the model studied. Indeed it has been recently demonstrated that it is possible to simulate SU($N$) impurity models with tunable exchange interactions reaching both FM and AFM regimes~\cite{Scazza:2014, riegger:2018}.

The iron pnictides display an intriguing QCP, as for example in BaFe$_2$(As$_{1-x}$P$_x$)$_2$~\cite{Jiang:2009,Kasahara:2010,walmsley:2013,Abrahams:2011}, where a divergent electron mass and concomitant destruction of the FL state was observed. This QCP has motivated several theoretical studies~\cite{fernandes:2013,levchenko:2013,chowdhury:2013}. 
Further progress from the perspective of this work would require the DMFT self-consistency condition and more realistic band structures.  In the DMFT treatment of a lattice model, the SFL and the NFL phases are expected to turn into magnetically ordered states, but the impurity model studied here with its power-law singularities would describe the behavior above $T_{\text{FL}}$.

The approach presented here, which takes into account the Hund's coupling and the multiorbital nature, is in the same spirit as the ideas of local quantum criticality used to describe Kondo breakdown using impurity models~\cite{Si2001}, so it would then be also useful for unconventional quantum phase transitions observed in other heavy-fermion materials~\cite{stewart:1984,takagi:1999,urano:2000,Gegenwart:2008}.

We thank H. Miao and R. Fernandes for helpful discussion. EW, KMS and JvD are supported by the Deutsche Forschungsgemeinschaft under Germany's Excellence Strategy--EXC-2111-390814868, and S-SBL by Grant No. LE3883/2-1. AW was funded by DOE-DE-SC0012704. YW and GK were supported by the US Department of energy, Office of Science, Basic Energy Sciences as a part of the Computational Materials Science Program through
the Center for Computational Design of Functional Strongly Correlated Materials and Theoretical Spectroscopy.

\bibliography{main}

\end{document}


\title{Supplemental Material for: \\ Global phase diagram of a spin-orbital Kondo impurity model and the suppression of Fermi-liquid scale}

\author{Y. Wang}     
\affiliation{Department of Condensed Matter Physics and Materials Science, Brookhaven National Laboratory, Upton, New York 11973, USA} 
\author{E. Walter}
\affiliation{Arnold Sommerfeld Center for Theoretical Physics, 
Center for NanoScience,\looseness=-1\,  and Munich 
Center for \\ Quantum Science and Technology,\looseness=-2\, Ludwig-Maximilians-Universit\"at M\"unchen, 80333 Munich, Germany}
\author{S.-S. B. Lee}
\affiliation{Arnold Sommerfeld Center for Theoretical Physics, 
Center for NanoScience,\looseness=-1\,  and Munich 
Center for \\ Quantum Science and Technology,\looseness=-2\, Ludwig-Maximilians-Universit\"at M\"unchen, 80333 Munich, Germany}
\author{K. M. Stadler}
\affiliation{Arnold Sommerfeld Center for Theoretical Physics, 
Center for NanoScience,\looseness=-1\,  and Munich 
Center for \\ Quantum Science and Technology,\looseness=-2\, Ludwig-Maximilians-Universit\"at M\"unchen, 80333 Munich, Germany}
\author{J. von Delft}
\affiliation{Arnold Sommerfeld Center for Theoretical Physics, 
Center for NanoScience,\looseness=-1\,  and Munich 
Center for \\ Quantum Science and Technology,\looseness=-2\, Ludwig-Maximilians-Universit\"at M\"unchen, 80333 Munich, Germany}
\author{A. Weichselbaum}
\affiliation{Department of Condensed Matter Physics and Materials Science, Brookhaven National Laboratory, Upton, New York 11973, USA} 
\affiliation{Arnold Sommerfeld Center for Theoretical Physics, 
Center for NanoScience,\looseness=-1\,  and Munich 
Center for \\ Quantum Science and Technology,\looseness=-2\, Ludwig-Maximilians-Universit\"at M\"unchen, 80333 Munich, Germany}
\author{G. Kotliar}
\affiliation{Department of Condensed Matter Physics and Materials Science, Brookhaven National Laboratory, Upton, New York 11973, USA} 
\affiliation{Department of Physics and Astronomy, Rutgers University, Piscataway, New Jersey 08856, USA}

\date{\today}
\maketitle

\section{NRG flow diagrams and dynamical susceptibilites}
\begin{figure*}
    \centering
    \includegraphics[width=0.98\textwidth]{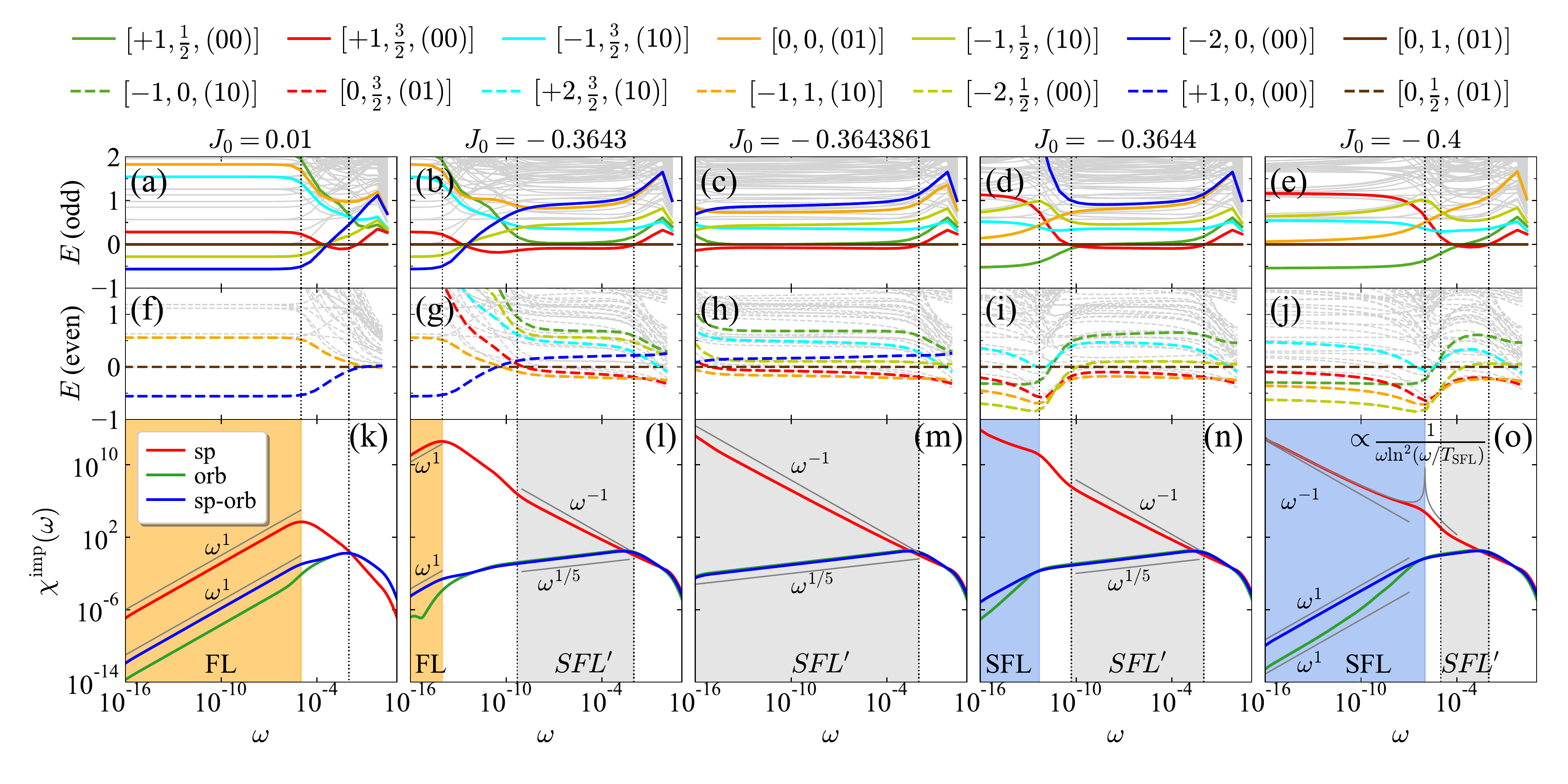}
    \caption{The phase transition from FL to SFL, with $I_0=0.01$.
         (a-j) Lin-log plots of NRG flow diagrams of Wilson chain with (a-e) odd and (f-j) even length $k$ as functions of the energy scale $\omega_{k}=\Lambda^{-k/2}$. The energy of multiplets $[0, 1, (01)]$ (brown solid)
         and $[0, \frac{1}{2}, (01)]$ (brown dashed)
         are chose as the reference energy $\mathcal{E}_{\text{ref}}$ for odd and even lengths, respectively. The symmetry labels of selected multiplets are shown at the top.
         (k-o) Log-log plots of the impurity dynamical susceptibilities $\chi^{\text{imp}}(\omega)$ for spin (red), orbital (green), and spin-orbital (blue) operators. }
    \vspace{1cm}

    \label{fig:flow1}
\end{figure*}

\begin{figure}
    \centering
    \includegraphics[width=0.48\textwidth]{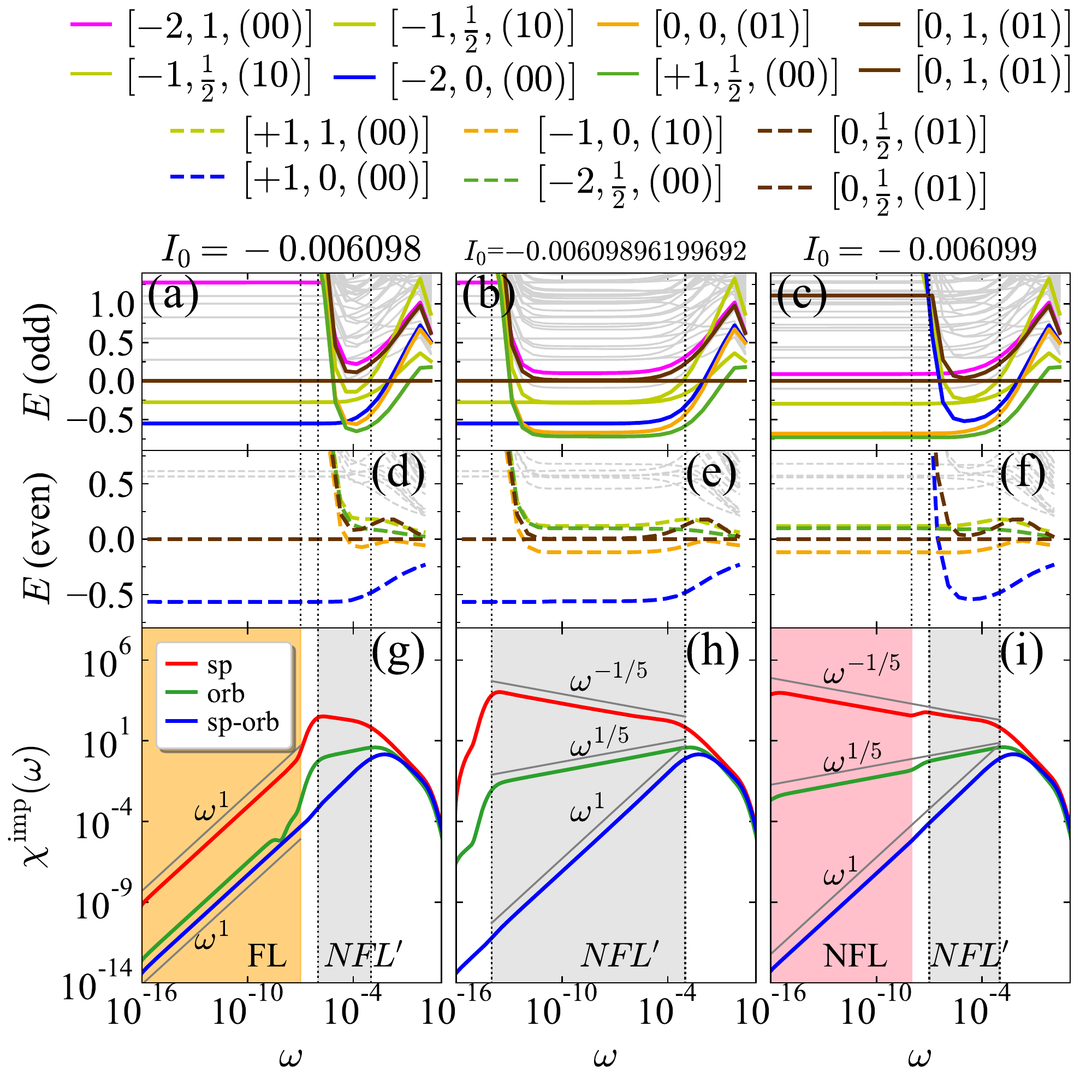}
    \caption{Analogous to Fig.~\ref{fig:flow1}, but for the phase transition from FL to NFL, with $J_0=0.3$.}
    \label{fig:flow2}
\end{figure}

\begin{figure}
    \centering
    \includegraphics[width=0.48\textwidth]{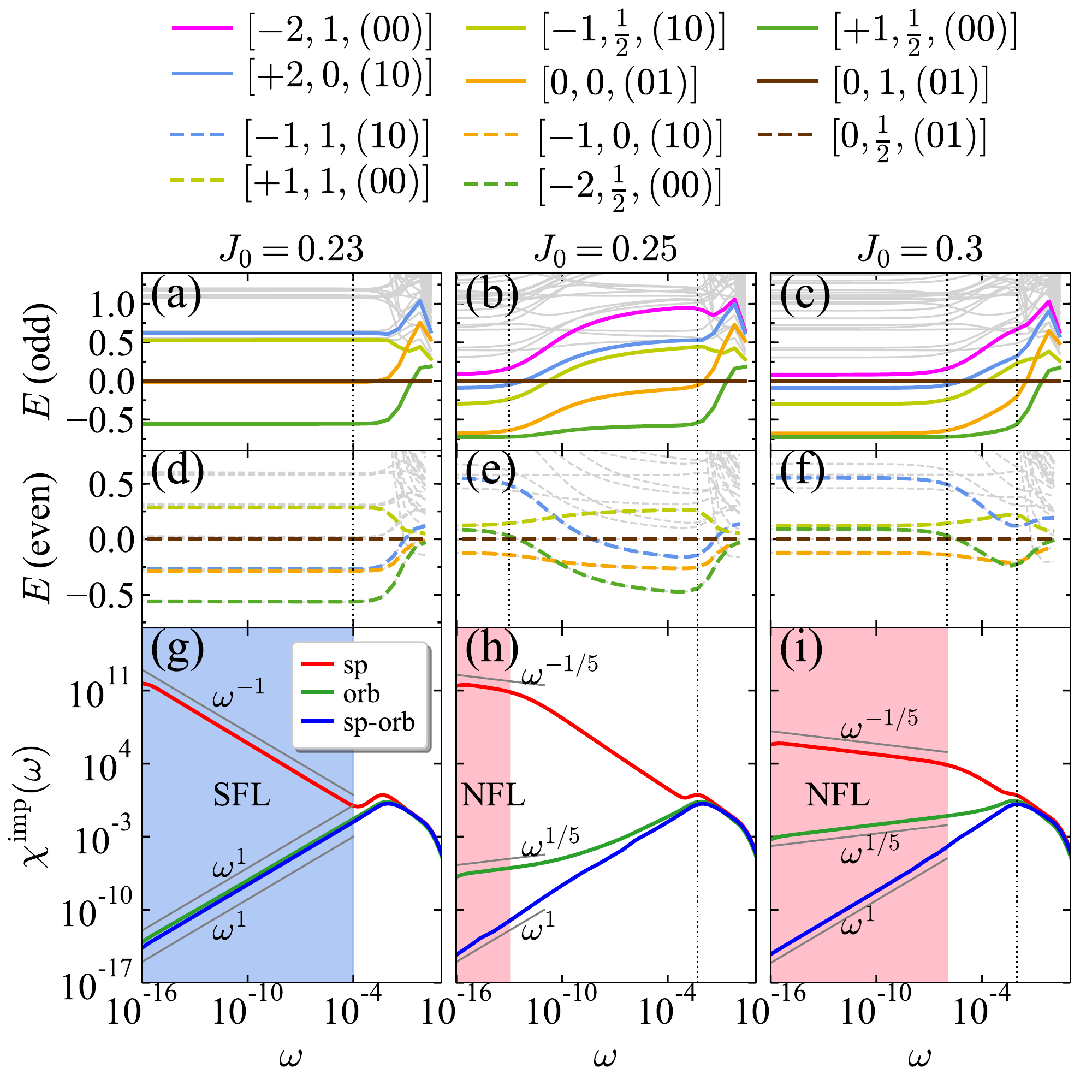}
    \caption{Analogous to Fig.~\ref{fig:flow1}, but for the phase transition from SFL to NFL, with $I_0=-0.01$.}
    \label{fig:flow3}
\end{figure}

In Figs.~\ref{fig:flow1}-\ref{fig:flow3}, we replot the NRG flow diagrams shown in the main text to add the flow diagrams
of Wilson chains with even length $k$. Here we follow the standard convention that $H_0$, the Hamiltonian of the impurity together with
the bath site at the location of the impurity, i.e., the bath site $k=0$, is a Wilson chain of even length. We use $K_0=0.3$, $\Lambda=4$ and half-bandwidth $D=1$, throughout.
In Fig.~\ref{fig:flow1}, we also add flow diagrams computed for $J_0=-0.3643861$ and $J_0=-0.3644$ to show more details of the phase transitions from FL to SFL. In Fig.~\ref{fig:compare_flow}, we compare the NRG flow diagrams of the NFL fixed point at $J_0=0.3, I_0=-0.01$ with those at $J_0=0.6, I_0=-0.13$.  As we move from the former to the latter, i.e. deeper into the NFL regimes, the eigenvalues of the fixed point spectrum change; this is caused by particle-hole asymmetry and can be explained by the CFT analysis [see Table~\ref{tab:cft3}] in Sec.~\ref{sec:cft}. 

The imaginary part of the dynamical susceptibilities of spin,
orbital and spin-orbital operators at the impurity site or the
zeroth bath site are defined as
\begin{subequations}
\begin{eqnarray}
    \chi^{\text{imp,bath}}_{\text{sp}}(\omega)&=&-\frac{1}{3\pi}\text{Im}\sum_{\alpha}\langle S^{\alpha}||S^{\alpha}\rangle_{\omega},\\ \chi^{\text{imp,bath}}_{\text{orb}}(\omega)&=&-\frac{1}{8\pi}\text{Im}\sum_{a}\langle T^{a}||T^a\rangle_{\omega},\\ \chi^{\text{imp,bath}}_{\text{sp-orb}}(\omega)&=&-\frac{1}{24\pi}\text{Im}\sum_{\alpha,a}\langle S^{\alpha} T^{a}||S^{\alpha} T^{a}\rangle_{\omega},
\end{eqnarray}
\end{subequations}
where, the operators $S^{\alpha}$ and $T^{a}$ refer to either the impurity site ($\chi^{\text{imp}}$) or the $k=0$ bath site ($\chi^{\text{bath}}$). The normalization averages over all correlators that are equivalent by the underlying symmetry. 

\begin{figure*}
    \centering
    \includegraphics[width=0.95\textwidth]{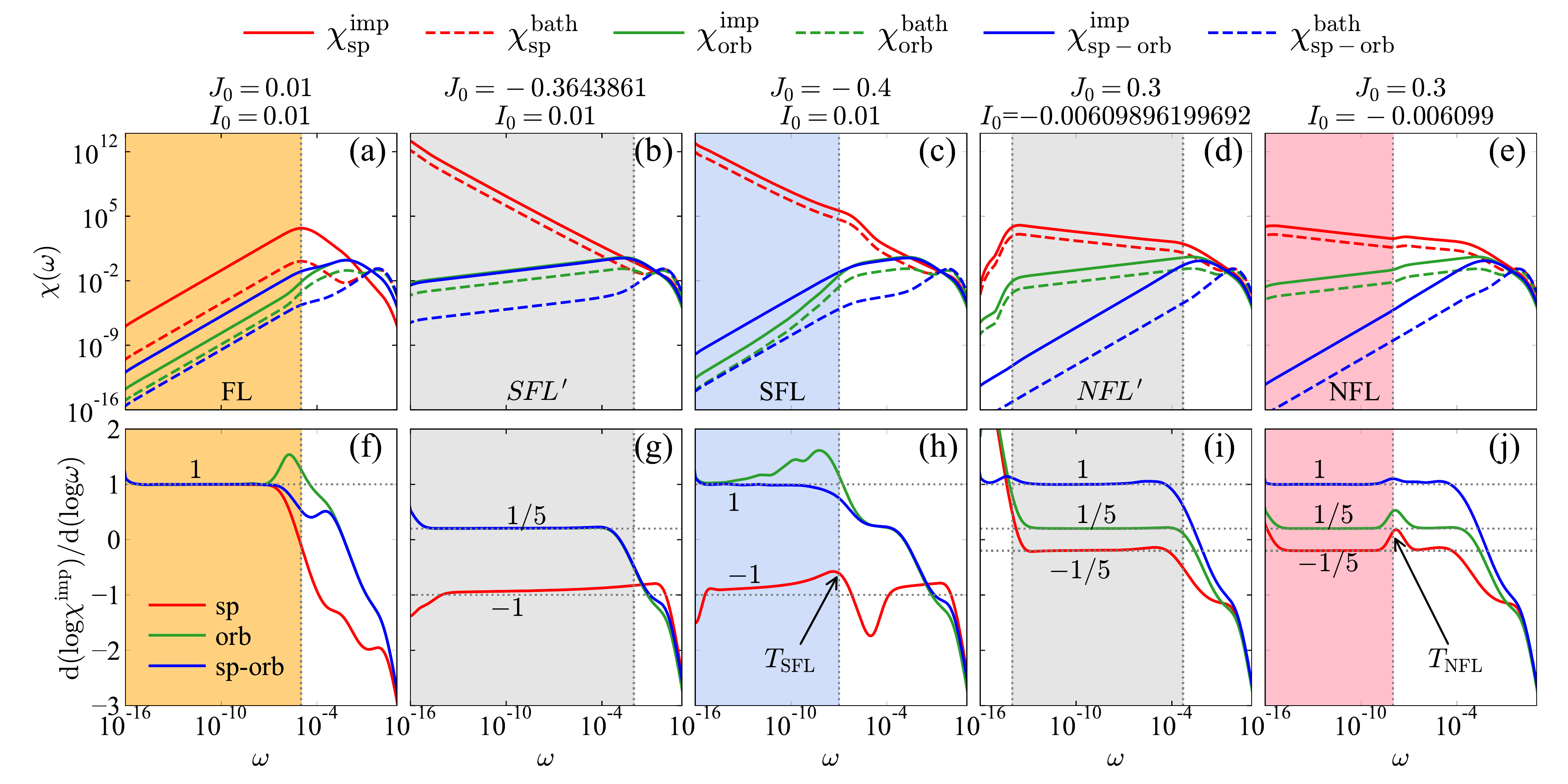}
    \caption{(a-e) Log-log plots of dynamical susceptibilities at
    impurity site $\chi^{\text{imp}}$ (solid) and the zeroth 
    bath site $\chi^{\text{bath}}$ (dashed). (f-j) Logarithmic derivative of the impurity dynamical susceptibilities,
$\frac{\text{d}(\text{log}\chi^{\text{imp}})}{\text{d}(\text{log}\omega)}$,
plotted using the same logarithmic $\omega$-axis as (a-e).
The numbers displayed near the curves give the corresponding (asymptotic) power-law exponents.
%
%
}
\label{fig:checkpower}
\end{figure*}

In Fig.~\ref{fig:checkpower}(a-e), we compare $\chi^{\text{imp}}(\omega)$ (solid lines) and $\chi^{\text{bath}}(\omega)$ (dashed lines). At the parameters we show, both follow the same behavior in the low-energy regimes of FL, SFL and NFL, and the intermediate-energy  regimes $\it{SFL}^{\prime}$ and $\it{NFL}^{\prime}$.

Figs.~\ref{fig:checkpower}(f-j) reveal the power laws governing $\chi^{\text{imp}}(\omega)$ shown in the main text and in Figs.~\ref{fig:flow1}-\ref{fig:flow3}, by showing the logarithmic derivative,
\begin{equation}
\label{eqn:power}
    \alpha(\omega)=\frac{\text{d}(\log\chi^{\text{imp}}(\omega))}{\text{d}(\log\omega)}.
\end{equation}
If $\chi^{\text{imp}}$ follows a pure power law, $\omega^{\alpha}$, its logarithmic derivative gives the constant exponent, $\alpha(\omega)=\alpha$. For FL, $\chi^{\text{imp}}(\omega)$ is linear at small $\omega$, $\alpha=1$, for all the three susceptibilities [Fig.~\ref{fig:checkpower}(f)], as expected for a FL.
For $\it{SFL}^{\prime}$ [Fig.~\ref{fig:checkpower}(g)], $\chi^{\text{imp}}_{\text{orb}}$ and $\chi^{\text{imp}}_{\text{sp-orb}}$ follow a well-defined power law with a constant value of $\alpha=1/5$. By contrast, $\chi^{\text{imp}}_{\text{sp}}$ does not quite, since $\alpha(\omega)$ shows slight $\omega$-dependence, indicating the presence of some non-power-law corrections. For SFL [Fig.~\ref{fig:checkpower}(h)], $\alpha(\omega)$ for both spin and orbital first increases  and then decreases, and finally approaches 
$-1$ and 1, respectively. This confirms the deviation from pure power-law behavior and the singular nature of this fixed point. However, $\chi^{\text{imp}}_{\text{sp-orb}}$ still follows a perfect power law. For $\it{NFL}^{\prime}$ [Figs.~\ref{fig:checkpower}(i)] and NFL [Figs.~\ref{fig:checkpower}(j)], $\chi^{\text{imp}}$ show well-defined power laws with $\alpha=-1/5$ for spin, $\alpha=1/5$ for orbital and $\alpha=1$ for spin-orbital, which are perfectly consistent with the CFT results presented in Sec.~\ref{sec:cft}.

\begin{figure*}
\includegraphics[width=0.95\textwidth]{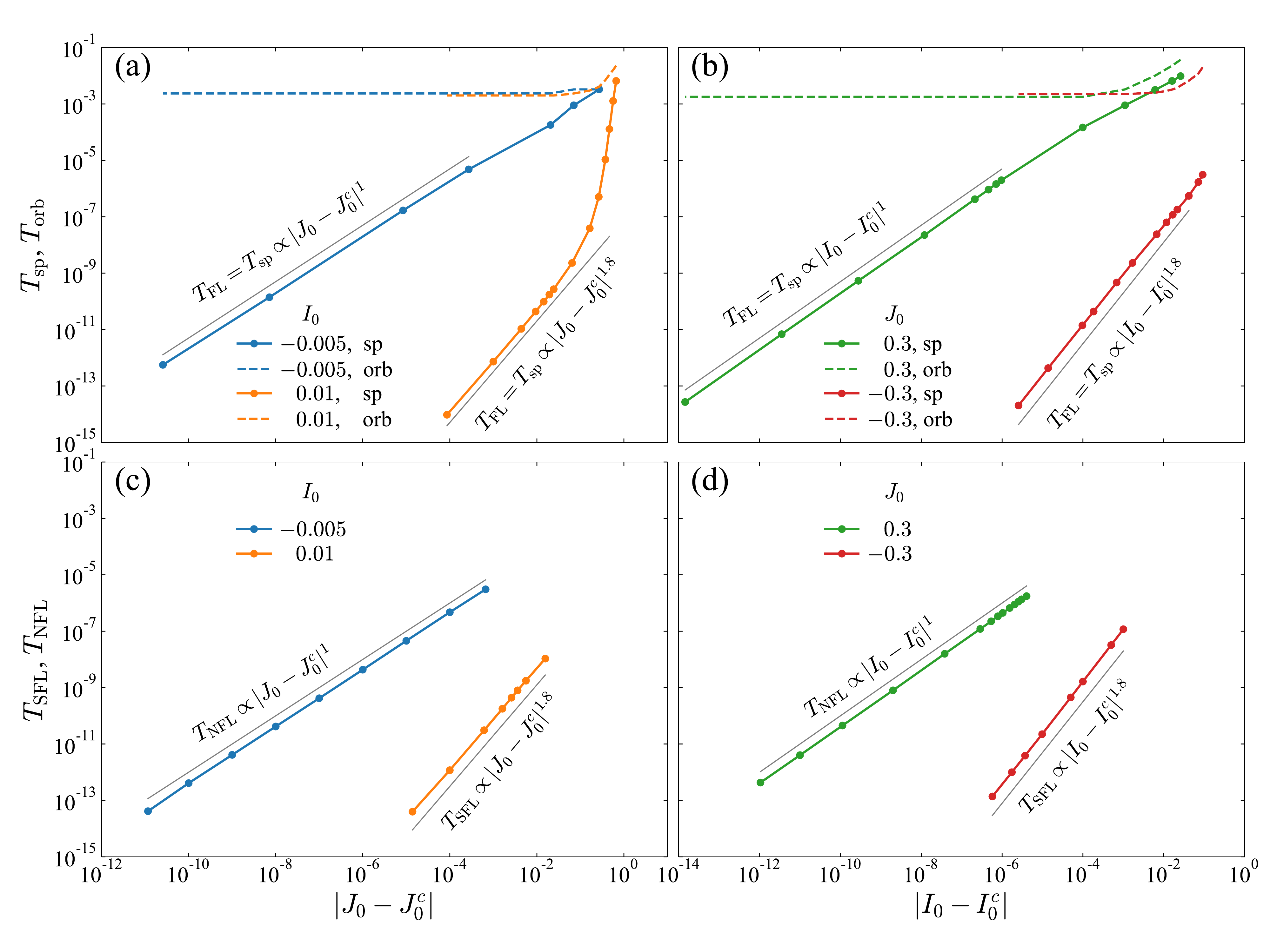}
\caption{Log-log plots of the spin and orbital Kondo scales, $T_{\text{sp}}$ (solid lines) and $T_{\text{orb}}$ (dashed lines), as functions of (a) $|J_0-J_0^{c}|$ for two values of $I_0$, or (b) $|I_0-I_0^{c}|$ for two values of $J_0$. Here, $T_{\text{sp}}$ can be considered as the FL scale, $T_{\text{FL}}=T_{\text{sp}}$, of the problem. The corresponding SFL and NFL scales, $T_{\text{SFL}}$ and $T_{\text{NFL}}$, as functions of (c) $|J_0-J_0^{c}|$, or (d) $|I_0-I_0^{c}|$.}
\label{fig:sos}
\end{figure*}

\begin{figure}
    \centering
    \includegraphics[width=0.48\textwidth]{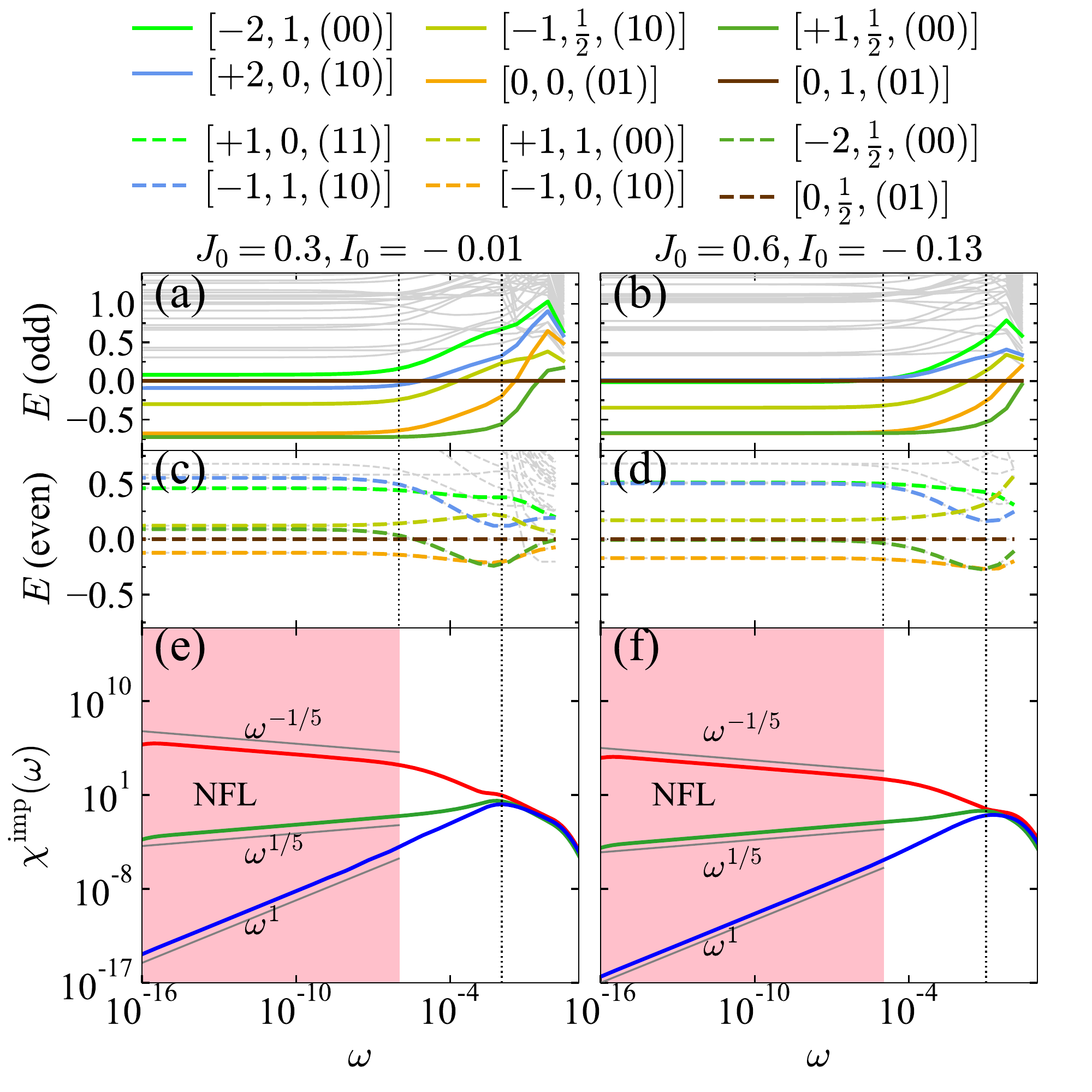}
    \caption{(a-d) NRG flow diagrams and (e-f) impurity dynamical susceptibilities, comparing the cases (a,c,e) $J_0=0.3, I_0=-0.01$ and (b,d,f) $J_0=0.6, I_0=-0.13$. (a,b) odd-$k$ and (c,d) even-$k$ Wilson chains.}
    \label{fig:compare_flow}
\end{figure}

\section{Destruction of Fermi-liquid scale}
We define the orbital (spin) Kondo scale $T_{\text{orb}}$ ($T_{\text{sp}}$) as the energy where $\chi^{\text{imp}}_{\text{orb (sp)}}$ is maximal. Fig.~\ref{fig:sos}(a) shows $T_{\text{sp}}$ (solid lines) and $T_{\text{orb}}$ (dashed lines) as functions of $|J_0-J_0^{c}|$ for two values of $I_0$, and Fig.~\ref{fig:sos}(b) shows them as functions of $|I_0-I_0^{c}|$ for two values of $J_0$, where $J_0^{c}$ and $I_0^{c}$ are the critical values at the phase transitions.
$T_{\text{orb}}$ remains large and almost constant throughout.
By contrast, $T_{\text{sp}}$ decreases in power-law fashion, $|J_0-J_0^{c}|^{\alpha}$ and $|I_0-I_0^{c}|^{\alpha}$, when approaching the phase boundary. Therefore, a spin-orbital separation (SOS) window~\cite{stadler:2015,stadler:2019} forms near the phase boundary, where $T_{\text{sp}}$ can be considered as the FL scale, $T_{\text{FL}}=T_{\text{sp}}$, of the problem. At the phase transitions, this FL scale vanishes, and a SFL or NFL scale arises. We define the energy scale characterizing the onset of SFL or NFL behavior [cf. Figs.1(b,c) in the main text], $T_{\text{SFL}}$ or $T_{\text{NFL}}$, as the energy where $\text{d}(\text{log}\chi^{\text{imp}}_{\text{sp}}(\omega))/\text{d}(\text{log}\omega)$ (Eq.~\ref{eqn:power}) is maximal, as indicated by arrows in Figs.~\ref{fig:checkpower}(h) and ~\ref{fig:checkpower}(j), respectively.
Fig.~\ref{fig:sos}(c) shows $T_{\text{SFL}}$ and $T_{\text{NFL}}$ as functions of $|J_0-J_0^{c}|$, and Fig.~\ref{fig:sos}(d) shows them as functions of $|I_0-I_0^{c}|$. $T_{\text{SFL}}$ or $T_{\text{NFL}}$ follows the same power-law behavior as $T_{\text{FL}}$ close to the phase boundary. The exponents found are $\alpha{=}1.8$ in the FL-SFL transition, and $\alpha{=}1$ for FL-NFL transition,
\begin{subequations}
\begin{eqnarray}
T_{\text{FL/SFL}}&\propto& |J_0-J_0^{c}|^{1.8}, \, |I_0-I_0^{c}|^{1.8},\\{}
T_{\text{FL/NFL}}&\propto& |J_0-J_0^{c}|^{1}, \;\;\, |I_0-I_0^{c}|^{1}.
\end{eqnarray}
\end{subequations}

\section{CFT analysis}
\label{sec:cft}
In this section, we follow the recently developed U(1)$\times$SU(2)$_3\times$SU(3)$_2$ conformal field theory (CFT) approach~\citep{walter:2019} to reproduce and understand the NRG finite-size spectra and the power laws of the dynamical susceptibilities
for the NFL fixed point. We start with a brief summary of the main results of Walter \textit{et al.}~\cite{walter:2019}, and refer the reader to that paper and references therein for more details on the CFT approach. The energy of the lowest multiplet, $Q\equiv [q, S, (\lambda_1\lambda_2)]$,
with charge $q$ relative to half-filling, spin $S$, and SU(3) orbital label in Dynkin notation $(\lambda_1 \lambda_2) \equiv (\lambda_1,\lambda_2)$, is given by
\begin{eqnarray}
E(Q; \delta q) 
&=&
\tfrac{1}{12}(q+\delta q)^2 +
\tfrac{1}{5} \kappa_2(S) 
+ \tfrac{1}{5} \kappa_3(\lambda_1, \lambda_2), 
\notag \\
    \kappa_2(S)&=&S(S+1), \label{eq:Eqdep}\\
      \kappa_3(\lambda_1,\lambda_2)&=&\tfrac{1}{3}(\lambda_1^2+\lambda_2^2+\lambda_1\lambda_2+3\lambda_1+3\lambda_2), \notag
\end{eqnarray}
where $\kappa_N$ represents 
the eigenvalues of the quadratic Casimir operator for SU($N$). 
$\delta q$ is a fitting parameter used to take into account particle-hole asymmetry effects, because the three-channel spin-orbital Kondo (3soK) model under consideration does not preserve particle-hole symmetry. Therefore $\delta q$
depends on the choice of the Kondo couplings $J_0$, $K_0$ and $I_0$. Its specific value can be determined via fits to the NRG finite-size spectra.
We simplify the formula by discarding the irrelevant constant term $\delta q^2/12$,
\begin{equation}
    E(Q,\delta q) 
    =
    \tfrac{1}{12} q^2 
   +\tfrac{\delta q}{6} q 
   +\tfrac{1}{5} \kappa_2(S) 
   +\tfrac{1}{5} \kappa_3(\lambda_1, \lambda_2). 
\end{equation}

In the main text, the impurity entropy in the NFL regime was found to be, $S_{\text{imp}}=\text{ln}\frac{1+\sqrt{5}}{2}$, [see Fig.\,1d]. This value can be obtained from either an 
SU(2)$_3$ Kondo model with overscreened spin, or an SU(3)$_2$ Kondo model with overscreened orbital. 
In the main text, we also observed that [see Figs.\,3(c) and 4(c)] the two lowest NRG eigenstates, $[+1,\frac{1}{2}, (00)]$ and $[0, 0, (01)]$, are very close in energy, and argued that the NFL phase undergoes a never-ending alternation of spin and orbital overscreening process, converting $[+1,\frac{1}{2}, (00)]$ to $[0, 0, (01)]$ and back [cf. Eqs. (2a,b) in the main text]. We take this as a motivation to apply either an
SU(2)$_3$ fusion procedure in the spin sector, or an SU(3)$_2$ fusion procedure in the orbital sector,
i.e. fusing the spectrum
of free fermions $Q\equiv [q, S, (\lambda_1\lambda_2)]$ with an effective impurity multiplet, $Q_{\text{imp}}^{\text{eff}}=[+1, \frac{1}{2}, (00)]$ or $Q_{\text{imp}}^{\text{eff}}=[0, 0, (01)]$,
to get the spectra of the NFL fixed point $Q'\equiv[q^\prime, S^\prime, (\lambda_1^{\prime}\lambda_2^{\prime})]$. The energy of the free fermions and the NFL spectra are given by $E(Q;0)$ and $E(Q';\delta q)$, respectively. 
Double fusion of the spectra $Q'$ with the conjugate, i.e., dual
representation of the effective impurity multiplet, $\bar{Q}_{\text{imp}}^{\text{eff}}=[-1, \frac{1}{2}, (00)]$ or $\bar{Q}_{\text{imp}}^{\text{eff}}=[0, 0, (10)]$, yields the quantum numbers $Q''\equiv [q^{\prime\prime}, S^{\prime\prime}, (\lambda_1^{\prime\prime}\lambda_2^{\prime\prime})]$, characterizing the CFT boundary operators with scaling dimensions given by $\Delta=E(Q'';\delta q)$.

Table~\ref{tab:cft1} and Table~\ref{tab:cft2} list the results of the two alternative fusion procedures with $Q_{\text{imp}}^{\text{eff}}=[+1, \frac{1}{2}, (00)]$ and $Q_{\text{imp}}^{\text{eff}}=[0, 0, (01)]$, respectively, and compare them with the energy $E_{\text{NRG}}$, computed by NRG for odd-$k$ Wilson chain at $J_0=0.3$ and $I_0=-0.01$. Remarkably, the two fusion procedures yield the \textit{same} fixed point spectrum, indicating that the SU(2)$_3$ and SU(3)$_2$ are actually equivalent and complementary descriptions of this NFL fixed point. Based on the CFT analysis, the energy difference of the mutiplets $[0, 0, (01)]$ and $[+1, \frac{1}{2}, (00)]$ is
\begin{equation}
    E([0, 0, (01)])-E([+1, \tfrac{1}{2}, (00)])=\frac{1-5\delta q}{30},
\end{equation}
hence $[+1, \frac{1}{2}, (00)]$ is the ground state when $\delta q < 0.2$, otherwise, $[0, 0, (01)]$ is the ground state.
Overall, we find for all of the explored region of the NFL phase that $|\delta q|\ll1$. Hence the energies above are nearly degenerate.
At $J_0=0.3$ and $I_0=-0.01$, the NRG calculation shows that $[+1, \frac{1}{2}, (00)]$ is the ground multiplet and $[0, 0, (01)]$ the first excited one. In Tables~\ref{tab:cft1}-\ref{tab:cft2}, $\delta E^{\prime}=E^{\prime}-E^{\prime}([+1,\frac{1}{2},(00)])$, is the excitation energy given by the CFT analysis. The NRG energies have been shifted and rescaled such that the ground state is zero and the values
of $E_{\text{NRG}}$ and $\delta E^{\prime}$ match for the multiplet $[+1,\frac{1}{2},(11)]$. Then $\delta q$ is determined by matching
$E_{\text{NRG}}$ and $\delta E^{\prime}$ for the multiplet $[0,0,(01)]$,
resulting in $\delta q\simeq 0.0433$. The remainder of the spectra
$\delta E^{\prime}$ and $E_{\text{NRG}}$ show
good 
agreement, with deviations smaller than $10\%$. 

In Table~\ref{tab:cft3}, we perform the same fusion procedure as that in Table~\ref{tab:cft2}, but compared with the NRG results at $J_0=0.6$ and $I_0=-0.13$, that is deep in the NFL phase diagram. The CFT can still perfectly reproduce the eigenlevel spectra in this region. It turns out that $\delta q \approx 0.2135$ is just above 0.2, such that the levels $[+1, \frac{1}{2}, (00)]$ and $[0,0,(01)]$ have crossed, with the latter the new ground state.
The remainder of the spectrum, which includes further level crossings, can also be perfectly reproduced by this value of $\delta q$.
The CFT analysis explains the change of the NRG flow diagrams
shown in Fig.~\ref{fig:compare_flow}.

With the double fusion procedure shown in Tables~\ref{tab:cft1}-\ref{tab:cft3}, the scaling dimension of spin, orbital and spin-orbital operators are found to be $\Delta_{\text{sp}}{=}\frac{2}{5}$, $\Delta_{\text{orb}}{=}\frac{3}{5}$ and $\Delta_{\text{sp-orb}}{=}1$, respectively, as highlighted by color in Tables~\ref{tab:cft1}-\ref{tab:cft3}. With this,
the power laws of the dynamical susceptibilites are predicted by CFT as~\cite{walter:2019}
\begin{subequations}
\begin{eqnarray}
    \chi^{\text{imp}}_{\text{sp}}\ \ &\sim& \omega^{2\Delta_{\text{sp}}-1}\ \ \ =\omega^{-1/5},\\ 
    \chi^{\text{imp}}_{\text{orb}}\ \ &\sim& \omega^{2\Delta_{\text{orb}}-1}\ \ =\omega^{1/5},\\ 
    \chi^{\text{imp}}_{\text{sp-orb}}&\sim& \omega^{2\Delta_{\text{sp-orb}}-1}=\omega^{1}.
\end{eqnarray}
\end{subequations}
in overall excellent agreement with the data in Figs.\,3(f) and 4(f) in the main text, and Figs.\,\ref{fig:flow2}(i),
\ref{fig:flow3}(i), \ref{fig:checkpower}(j), \ref{fig:compare_flow}(e,f) in this Supplemental Material.

\begin{table*}
\caption{Fusion table for the NFL fixed point.
Left: The 12 lowest low-lying free
fermion multiplets $Q=[q,S,(\lambda_{1}\lambda_{2})]$ with
degeneracy $d$ and energies $E(Q;0)$. Middle:
``Single fusion'' with an effective impurity multiplet
$Q_\mathrm{imp}^{\mathrm{eff}}=[+1,\frac{1}{2},(00)]$,
using $\text{SU(2)}_{3}$ fusion rules~\cite{walter:2019} in the spin sector.
This yields multiplets $Q^{\prime}=[q^{\prime},S^{\prime},(\lambda_{1}^{\prime}\lambda_{2}^{\prime})]$ with degeneracy $d^{\prime}$ and energies $E^{\prime}=E(Q^{\prime};\delta q)$.
The excitation energies are $\delta E^{\prime}=E^{\prime}-E_{\text{min}}^{\prime}$, where $E_{\text{min}}^{\prime}=E^{\prime}([+1,\frac{1}{2},(00)])=\frac{7+5\delta q}{30}$.
These are compared to the values $E_{\text{NRG}}$, computed by NRG
at $J_{0}=0.3,K_{0}=0.3,I_{0}=-0.01$, where $[+1, \frac{1}{2}, (00)]$ is the ground state. The NRG energies have been
shifted and rescaled such that the ground state is zero and the values
of $E_{\text{NRG}}$ and $\delta E^{\prime}$ match for the multiplet
$[+1,\frac{1}{2},(11)]$. $\delta q$ is then determined by matching
$E_{\text{NRG}}$ and $\delta E^{\prime}$ for the multiplet $[0,0,(01)]$,
resulting in $\delta q=0.0433$. Right: ``Double fusion'', which fuses
multiplets from the middle column with an impurity in the conjugate
representation $\bar{Q}_{\text{imp}}^{\text{eff}}=[-1,\frac{1}{2},(00)]$, yields
the quantum numbers $Q^{\prime\prime}=[q^{\prime\prime},S^{\prime\prime},(\lambda_{1}^{\prime\prime}\lambda_{2}^{\prime\prime})]$.
These characterize the CFT boundary operators $\hat{O}$, with scaling
dimensions $\Delta=E(Q^{\prime\prime};\delta q)$.
}
\label{tab:cft1}

\renewcommand{\arraystretch}{1.6}

\begin{tabular}{ccccc|cccccc|c|ccccc}

\hline 
\hline
\multicolumn{5}{c|}{free fermions} & \multicolumn{6}{c|}{single fusion, with $Q_{\text{imp}}^{\text{eff}}=[+1,\frac{1}{2},(00)]$} & NRG & \multicolumn{5}{c}{double fusion, with $\bar{Q}_{\text{imp}}^{\text{eff}}=[-1,\frac{1}{2},(00)]$}\tabularnewline[\doublerulesep]
$q$ & $S$ & $(\lambda_{1}\lambda_{2})$ & $d$ & $E$ & $q^{\prime}$ & $S^{\prime}$ & $(\lambda_{1}^{\prime}\lambda_{2}^{\prime})$ & $d^{\prime}$ & $E^{\prime}$ & $\delta E^{\prime}$ & $E_{\text{NRG}}$ & $q^{\prime\prime}$ & $S^{\prime\prime}$ & $(\lambda_{1}^{\prime\prime}\lambda_{2}^{\prime\prime})$ & $\Delta$ & $\hat{O}$\tabularnewline
\hline 
\hline 
\multirow{2}{*}{0} & \multirow{2}{*}{0} & \multirow{2}{*}{(00)} & \multirow{2}{*}{1} & \multirow{2}{*}{0} & \multirow{2}{*}{+1} & \multirow{2}{*}{$\frac{1}{2}$} & \multirow{2}{*}{(00)} & \multirow{2}{*}{2} & \multirow{2}{*}{$\frac{7+5\delta q}{30}$} & \multirow{2}{*}{0} & \multirow{2}{*}{0} & 0 & 0 & (00) & 0 & $\mathds{1}$\tabularnewline
\cline{13-17} 
 &  &  &  &  &  &  &  &  &  &  &  & 0 & 1 & (00) & \textcolor{red}{$\frac{2}{5}$ ($=\Delta_{\text{sp}}$)}& $\Phi_{\text{sp}}$\tabularnewline
\hline 
\multirow{3}{*}{+1} & \multirow{3}{*}{$\frac{1}{2}$} & \multirow{3}{*}{(10)} & \multirow{3}{*}{6} & \multirow{3}{*}{$\frac{1}{2}$} & +2 & 0 & (10) & 3 & $\frac{9+5\delta q}{15}$ & $\frac{11+5\delta q}{30}$ (0.374) & 0.369 & +1 & $\frac{1}{2}$ & (10) & $\frac{3+\delta q}{6}$ & \tabularnewline
\cline{6-17} 
 &  &  &  &  & \multirow{2}{*}{+2} & \multirow{2}{*}{1} & \multirow{2}{*}{(10)} & \multirow{2}{*}{9} & \multirow{2}{*}{$\frac{3+\delta q}{3}$} & \multirow{2}{*}{$\frac{23+5\delta q}{30}$ (0.774)} & \multirow{2}{*}{0.809} & +1 & $\frac{1}{2}$ & (10) & $\frac{3+\delta q}{6}$ & \tabularnewline
\cline{13-17} 
 &  &  &  &  &  &  &  &  &  &  &  & +1 & $\frac{3}{2}$ & (10) & $\frac{33+5\delta q}{30}$ & \tabularnewline
\hline 
\multirow{3}{*}{$-1$} & \multirow{3}{*}{$\frac{1}{2}$} & \multirow{3}{*}{$(01)$} & \multirow{3}{*}{6} & \multirow{3}{*}{$\frac{1}{2}$} & 0 & 0 & (01) & 3 & $\frac{4}{15}$ & $\frac{1-5\delta q}{30}$ (0.026) & 0.026 & $-1$ & $\frac{1}{2}$ & (01) & $\frac{3-\delta q}{6}$ & \tabularnewline
\cline{6-17} 
 &  &  &  &  & \multirow{2}{*}{0} & \multirow{2}{*}{1} & \multirow{2}{*}{(01)} & \multirow{2}{*}{9} & \multirow{2}{*}{$\frac{2}{3}$} & \multirow{2}{*}{$\frac{13-5\delta q}{30}$ (0.426)} & \multirow{2}{*}{0.422} & $-1$ & $\frac{1}{2}$ & (01) & $\frac{3-\delta q}{6}$ & \tabularnewline
\cline{13-17} 
 &  &  &  &  &  &  &  &  &  &  &  & $-1$ & $\frac{3}{2}$ & (01) & $\frac{33-5\delta q}{30}$ & \tabularnewline
\hline 
\multirow{3}{*}{0} & \multirow{3}{*}{1} & \multirow{3}{*}{(11)} & \multirow{3}{*}{24} & \multirow{3}{*}{1} & \multirow{2}{*}{+1} & \multirow{2}{*}{$\frac{1}{2}$} & \multirow{2}{*}{(11)} & \multirow{2}{*}{16} & \multirow{2}{*}{$\frac{5+\delta q}{6}$} & \multirow{2}{*}{$\frac{3}{5}$ (0.600)} & \multirow{2}{*}{0.600} & 0 & 0 & (11) & \textcolor{ForestGreen}{$\frac{3}{5}$ ($=\Delta_{\text{orb}}$)} & $\Phi_{\text{orb}}$\tabularnewline
\cline{13-17} 
 &  &  &  &  &  &  &  &  &  &  &  & 0 & 1 & (11) & \textcolor{blue}{1 ($=\Delta_{\text{sp-orb}}$)} & $\Phi_{\text{sp-orb}}$\tabularnewline
\cline{6-17} 
 &  &  &  &  & +1 & $\frac{3}{2}$ & (11) & 32 & $\frac{43+5\delta q}{30}$ & $\frac{6}{5}$ (1.200) & 1.223 & 0 & 1 & (11) & 1 & \tabularnewline
\hline 
\multirow{2}{*}{+2} & \multirow{2}{*}{0} & \multirow{2}{*}{(20)} & \multirow{2}{*}{6} & \multirow{2}{*}{1} & \multirow{2}{*}{+3} & \multirow{2}{*}{$\frac{1}{2}$} & \multirow{2}{*}{(20)} & \multirow{2}{*}{12} & \multirow{2}{*}{$\frac{47+15\delta q}{30}$} & \multirow{2}{*}{$\frac{4+\delta q}{3}$ (1.348)} & \multirow{2}{*}{1.432} & +2 & 0 & (20) & $\frac{3+\delta q}{3}$ & \tabularnewline
\cline{13-17} 
 &  &  &  &  &  &  &  &  &  &  &  & +2 & 1 & (20) & $\frac{21+5\delta q}{15}$ & \tabularnewline
\hline 
\multirow{2}{*}{$-2$} & \multirow{2}{*}{0} & \multirow{2}{*}{(02)} & \multirow{2}{*}{6} & \multirow{2}{*}{1} & \multirow{2}{*}{$-1$} & \multirow{2}{*}{$\frac{1}{2}$} & \multirow{2}{*}{(02)} & \multirow{2}{*}{12} & \multirow{2}{*}{$\frac{27-5\delta q}{30}$} & \multirow{2}{*}{$\frac{2-\delta q}{3}$ (0.652)} & \multirow{2}{*}{0.655} & $-2$ & 0 & (02) & $\frac{3-\delta q}{3}$ & \tabularnewline
\cline{13-17} 
 &  &  &  &  &  &  &  &  &  &  &  & $-2$ & 1 & (02) & $\frac{21-5\delta q}{15}$ & \tabularnewline
\hline 
\multirow{3}{*}{+2} & \multirow{3}{*}{1} & \multirow{3}{*}{(01)} & \multirow{3}{*}{9} & \multirow{3}{*}{1} & \multirow{2}{*}{+3} & \multirow{2}{*}{$\frac{1}{2}$} & \multirow{2}{*}{(01)} & \multirow{2}{*}{6} & \multirow{2}{*}{$\frac{7+3\delta q}{6}$} & \multirow{2}{*}{$\frac{14+5\delta q}{15}$ (0.948)} & \multirow{2}{*}{0.954} & +2 & 0 & (01) & $\frac{9+5\delta q}{15}$ & \tabularnewline
\cline{13-17} 
 &  &  &  &  &  &  &  &  &  &  &  & +2 & 1 & (01) & $\frac{3+\delta q}{3}$ & \tabularnewline
\cline{6-17} 
 &  &  &  &  & +3 & $\frac{3}{2}$ & (01) & 12 & $\frac{53+15\delta q}{30}$ & $\frac{23+5\delta q}{15}$ (1.548) & 1.599 & +2 & 1 & (01) & $\frac{3+\delta q}{3}$ & \tabularnewline
\hline 
\multirow{3}{*}{$-2$} & \multirow{3}{*}{1} & \multirow{3}{*}{(10)} & \multirow{3}{*}{9} & \multirow{3}{*}{1} & \multirow{2}{*}{$-1$} & \multirow{2}{*}{$\frac{1}{2}$} & \multirow{2}{*}{(10)} & \multirow{2}{*}{6} & \multirow{2}{*}{$\frac{3-\delta q}{6}$} & \multirow{2}{*}{$\frac{4-5\delta q}{15}$ (0.252)} & \multirow{2}{*}{0.248} & $-2$ & 0 & (10) & $\frac{9-5\delta q}{15}$ & \tabularnewline
\cline{13-17} 
 &  &  &  &  &  &  &  &  &  &  &  & $-2$ & 1 & (10) & $\frac{3-\delta q}{3}$ & \tabularnewline
\cline{6-17} 
 &  &  &  &  & $-1$ & $\frac{3}{2}$ & (10) & 12 & $\frac{33-5\delta q}{30}$ & $\frac{13-5\delta q}{15}$ (0.852) & 0.844 & $-2$ & 1 & (10) & $\frac{3-\delta q}{3}$ & \tabularnewline
\hline 
\multirow{2}{*}{+1} & \multirow{2}{*}{$\frac{3}{2}$} & \multirow{2}{*}{(02)} & \multirow{2}{*}{24} & \multirow{2}{*}{$\frac{3}{2}$} & \multirow{2}{*}{+2} & \multirow{2}{*}{1} & \multirow{2}{*}{(02)} & \multirow{2}{*}{18} & \multirow{2}{*}{$\frac{21+5\delta q}{15}$} & \multirow{2}{*}{$\frac{7+\delta q}{6}$ (1.174)} & \multirow{2}{*}{1.180} & +1 & $\frac{1}{2}$ & (02) & $\frac{27+5\delta q}{30}$ & \tabularnewline
\cline{13-17} 
 &  &  &  &  &  &  &  &  &  &  &  & +1 & $\frac{3}{2}$ & (02) & $\frac{9+\delta q}{6}$ & \tabularnewline
\hline 
\multirow{2}{*}{$-1$} & \multirow{2}{*}{$\frac{3}{2}$} & \multirow{2}{*}{(20)} & \multirow{2}{*}{24} & \multirow{2}{*}{$\frac{3}{2}$} & \multirow{2}{*}{0} & \multirow{2}{*}{1} & \multirow{2}{*}{(20)} & \multirow{2}{*}{18} & \multirow{2}{*}{$\frac{16}{15}$} & \multirow{2}{*}{$\frac{5-\delta q}{6}$ (0.826)} & \multirow{2}{*}{0.825} & $-1$ & $\frac{1}{2}$ & (20) & $\frac{27-5\delta q}{30}$ & \tabularnewline
\cline{13-17} 
 &  &  &  &  &  &  &  &  &  &  &  & $-1$ & $\frac{3}{2}$ & (20) & $\frac{9-\delta q}{6}$ & \tabularnewline
\hline 
\multirow{3}{*}{$-3$} & \multirow{3}{*}{$\frac{1}{2}$} & \multirow{3}{*}{(11)} & \multirow{3}{*}{16} & \multirow{3}{*}{$\frac{3}{2}$} & $-2$ & 0 & (11) & 8 & $\frac{14-5\delta q}{15}$ & $\frac{7-5\delta q}{10}$ (0.678) & 0.673 & $-3$ & $\frac{1}{2}$ & (11) & $\frac{3-\delta q}{2}$ & \tabularnewline
\cline{6-17} 
 &  &  &  &  & \multirow{2}{*}{$-2$} & \multirow{2}{*}{1} & \multirow{2}{*}{(11)} & \multirow{2}{*}{24} & \multirow{2}{*}{$\frac{4-\delta q}{3}$} & \multirow{2}{*}{$\frac{11-5\delta q}{10}$ (1.078)} & \multirow{2}{*}{1.090} & $-3$ & $\frac{1}{2}$ & (11) & $\frac{3-\delta q}{2}$ & \tabularnewline
\cline{13-17} 
 &  &  &  &  &  &  &  &  &  &  &  & $-3$ & $\frac{3}{2}$ & (11) & $\frac{21-5\delta q}{10}$ & \tabularnewline
\hline 
\multirow{2}{*}{$-3$} & \multirow{2}{*}{$\frac{3}{2}$} & \multirow{2}{*}{(00)} & \multirow{2}{*}{4} & \multirow{2}{*}{$\frac{3}{2}$} & \multirow{2}{*}{$-2$} & \multirow{2}{*}{1} & \multirow{2}{*}{(00)} & \multirow{2}{*}{3} & \multirow{2}{*}{$\frac{11-5\delta q}{15}$} & \multirow{2}{*}{$\frac{1-\delta q}{2}$ (0.478)} & \multirow{2}{*}{0.470} & $-3$ & $\frac{1}{2}$ & (00) & $\frac{9-5\delta q}{10}$ & \tabularnewline
 &  &  &  &  &  &  &  &  &  &  &  & $-3$ & $\frac{3}{2}$ & (00) & $\frac{3-\delta q}{2}$ & \tabularnewline
 \hline
 \hline
\end{tabular}
\end{table*}

\begin{table*}
\caption{Fusion table for the NFL fixed point constructed in analogous to Table~\ref{tab:cft1}, but now using $\text{SU(3)}_{2}$ fusion rules~\cite{walter:2019} in the orbital sector, with $Q_\mathrm{imp}^{\text{eff}}=[0,0,(01)]$ for single fusion and $\bar{Q}_{\text{imp}}^{\text{eff}}=[0,0,(10)]$ for double fusion.
The excitation energies are $\delta E^{\prime}=E^{\prime}-E_{\text{min}}^{\prime}$, where $E_{\text{min}}^{\prime}=E^{\prime}([+1,\frac{1}{2},(00)])=\frac{7+5\delta q}{30}$.
These are compared to the values $E_{\text{NRG}}$, computed by NRG
at $J_{0}=0.3,K_{0}=0.3,I_{0}=-0.01$, where $[+1, \frac{1}{2}, (00)]$ is the ground state (same as for Table~\ref{tab:cft1}). 
The values of $E_{\text{NRG}}$ and $\delta E^{\prime}$ match for the multiplet
$[+1,\frac{1}{2},(11)]$. $\delta q$ is then determined by matching
$E_{\text{NRG}}$ and $\delta E^{\prime}$ for the multiplet $[0,0,(01)]$,
resulting in $\delta q=0.0433$.}
\label{tab:cft2}

\renewcommand{\arraystretch}{1.6}

\begin{tabular}{ccccc|cccccc|c|ccccc}
\hline
\hline
\multicolumn{5}{c|}{free fermions} & \multicolumn{6}{c|}{single fusion, with $Q_{\text{imp}}^{\text{eff}}=[0,0,(01)]$} & NRG & \multicolumn{5}{c}{double fusion, with $\bar{Q}_{\text{imp}}^{\text{eff}}=[0,0,(10)]$}\tabularnewline
$q$ & $S$ & $(\lambda_{1}\lambda_{2})$ & $d$ & $E$ & $q^{\prime}$ & $S^{\prime}$ & $(\lambda_{1}^{\prime}\lambda_{2}^{\prime})$ & $d^{\prime}$ & $E^{\prime}$ & $\delta E^{\prime}$ & $E_{\text{NRG}}$ & $q^{\prime\prime}$ & $S^{\prime\prime}$ & $(\lambda_{1}^{\prime\prime}\lambda_{2}^{\prime\prime})$ & $\Delta$ & $\hat{O}$\tabularnewline
\hline 
\hline
\multirow{2}{*}{0} & \multirow{2}{*}{0} & \multirow{2}{*}{(00)} & \multirow{2}{*}{1} & \multirow{2}{*}{0} & \multirow{2}{*}{0} & \multirow{2}{*}{0} & \multirow{2}{*}{(01)} & \multirow{2}{*}{3} & \multirow{2}{*}{$\frac{4}{15}$} & \multirow{2}{*}{$\frac{1-5\delta q}{30}$ (0.026)} & \multirow{2}{*}{0.026} & 0 & 0 & (00) & 0 & $\mathds{1}$\tabularnewline
\cline{13-17} 
 &  &  &  &  &  &  &  &  &  &  &  & 0 & 0 & (11) & \textcolor{ForestGreen}{$\frac{3}{5}$ ($=\Delta_{\text{orb}}$)} & $\Phi_{\text{orb}}$\tabularnewline
\hline 
\multirow{3}{*}{+1} & \multirow{3}{*}{$\frac{1}{2}$} & \multirow{3}{*}{(10)} & \multirow{3}{*}{6} & \multirow{3}{*}{$\frac{1}{2}$} & +1 & $\frac{1}{2}$ & (00) & 2 & $\frac{7+5\delta q}{30}$ & 0 & 0 & +1 & $\frac{1}{2}$ & (10) & $\frac{3+\delta q}{6}$ & \tabularnewline
\cline{6-17} 
 &  &  &  &  & \multirow{2}{*}{+1} & \multirow{2}{*}{$\frac{1}{2}$} & \multirow{2}{*}{(11)} & \multirow{2}{*}{16} & \multirow{2}{*}{$\frac{5+\delta q}{6}$} & \multirow{2}{*}{$\frac{3}{5}$ (0.600)} & \multirow{2}{*}{0.600} & +1 & $\frac{1}{2}$ & (10) & $\frac{3+\delta q}{6}$ & \tabularnewline
\cline{13-17} 
 &  &  &  &  &  &  &  &  &  &  &  & +1 & $\frac{1}{2}$ & (02) & $\frac{27+5\delta q}{30}$ & \tabularnewline
\hline 
\multirow{3}{*}{$-1$} & \multirow{3}{*}{$\frac{1}{2}$} & \multirow{3}{*}{(01)} & \multirow{3}{*}{6} & \multirow{3}{*}{$\frac{1}{2}$} & \multirow{2}{*}{$-1$} & \multirow{2}{*}{$\frac{1}{2}$} & \multirow{2}{*}{(10)} & \multirow{2}{*}{6} & \multirow{2}{*}{$\frac{3-\delta q}{6}$} & \multirow{2}{*}{$\frac{4-5\delta q}{15}$ (0.252)} & \multirow{2}{*}{0.248} & $-1$ & $\frac{1}{2}$ & (01) & $\frac{3-\delta q}{6}$ & \tabularnewline
\cline{13-17} 
 &  &  &  &  &  &  &  &  &  &  &  & $-1$ & $\frac{1}{2}$ & (20) & $\frac{27-5\delta q}{30}$ & \tabularnewline
\cline{6-17} 
 &  &  &  &  & $-1$ & $\frac{1}{2}$ & (02) & 12 & $\frac{27-5\delta q}{30}$ & $\frac{2-\delta q}{3}$ (0.652) & 0.655 & $-1$ & $\frac{1}{2}$ & (01) & $\frac{3-\delta q}{6}$ & \tabularnewline
\hline 
\multirow{3}{*}{0} & \multirow{3}{*}{1} & \multirow{3}{*}{(11)} & \multirow{3}{*}{24} & \multirow{3}{*}{1} & \multirow{2}{*}{0} & \multirow{2}{*}{1} & \multirow{2}{*}{(01)} & \multirow{2}{*}{9} & \multirow{2}{*}{$\frac{2}{3}$} & \multirow{2}{*}{$\frac{13-5\delta q}{30}$ (0.426)} & \multirow{2}{*}{0.422} & 0 & 1 & (00) & \textcolor{red}{$\frac{2}{5}$ ($=\Delta_{\text{sp}}$)} & $\Phi_{\text{sp}}$\tabularnewline
\cline{13-17} 
 &  &  &  &  &  &  &  &  &  &  &  & 0 & 1 & (11) & \textcolor{blue}{1 ($=\Delta_{\text{sp-orb}}$)} & $\Phi_{\text{sp-orb}}$ \tabularnewline
\cline{6-17} 
 &  &  &  &  & 0 & 1 & (20) & 18 & $\frac{16}{15}$ & $\frac{5-\delta q}{6}$ (0.826) & 0.825 & 0 & 1 & (11) & 1 & \tabularnewline
\hline 
\multirow{2}{*}{+2} & \multirow{2}{*}{0} & \multirow{2}{*}{(20)} & \multirow{2}{*}{6} & \multirow{2}{*}{1} & \multirow{2}{*}{+2} & \multirow{2}{*}{0} & \multirow{2}{*}{(10)} & \multirow{2}{*}{3} & \multirow{2}{*}{$\frac{9+5\delta q}{15}$} & \multirow{2}{*}{$\frac{11+5\delta q}{30}$ (0.374)} & \multirow{2}{*}{0.369} & +2 & 0 & (01) & $\frac{9+5\delta q}{15}$ & \tabularnewline
\cline{13-17} 
 &  &  &  &  &  &  &  &  &  &  &  & +2 & 0 & (20) & $\frac{3+\delta q}{3}$ & \tabularnewline
\hline 
\multirow{2}{*}{$-2$} & \multirow{2}{*}{0} & \multirow{2}{*}{(02)} & \multirow{2}{*}{6} & \multirow{2}{*}{1} & \multirow{2}{*}{$-2$} & \multirow{2}{*}{0} & \multirow{2}{*}{(11)} & \multirow{2}{*}{8} & \multirow{2}{*}{$\frac{14-5\delta q}{15}$} & \multirow{2}{*}{$\frac{7-5\delta q}{10}$ (0.678)} & \multirow{2}{*}{0.673} & $-2$ & 0 & (10) & $\frac{9-5\delta q}{15}$ & \tabularnewline
\cline{13-17} 
 &  &  &  &  &  &  &  &  &  &  &  & $-2$ & 0 & (02) & $\frac{3-\delta q}{3}$ & \tabularnewline
\hline 
\multirow{3}{*}{+2} & \multirow{3}{*}{1} & \multirow{3}{*}{(01)} & \multirow{3}{*}{9} & \multirow{3}{*}{1} & \multirow{2}{*}{+2} & \multirow{2}{*}{1} & \multirow{2}{*}{(10)} & \multirow{2}{*}{9} & \multirow{2}{*}{$\frac{3+\delta q}{3}$} & \multirow{2}{*}{$\frac{23+5\delta q}{30}$ (0.774)} & \multirow{2}{*}{0.809} & +2 & 1 & (01) & $\frac{3+\delta q}{3}$ & \tabularnewline
\cline{13-17} 
 &  &  &  &  &  &  &  &  &  &  &  & +2 & 1 & (20) & $\frac{21+5\delta q}{15}$ & \tabularnewline
\cline{6-17} 
 &  &  &  &  & +2 & 1 & (02) & 18 & $\frac{21+5\delta q}{15}$ & $\frac{7+\delta q}{6}$ (1.174) & 1.180 & +2 & 1 & (01) & $\frac{3+\delta q}{3}$ & \tabularnewline
\hline 
\multirow{3}{*}{$-2$} & \multirow{3}{*}{1} & \multirow{3}{*}{(10)} & \multirow{3}{*}{9} & \multirow{3}{*}{1} & $-2$ & 1 & (00) & 3 & $\frac{11-5\delta q}{15}$ & $\frac{1-\delta q}{2}$ (0.478) & 0.470 & $-2$ & 1 & (10) & $\frac{3-\delta q}{3}$ & \tabularnewline
\cline{6-17} 
 &  &  &  &  & \multirow{2}{*}{$-2$} & \multirow{2}{*}{1} & \multirow{2}{*}{(11)} & \multirow{2}{*}{24} & \multirow{2}{*}{$\frac{4-\delta q}{3}$} & \multirow{2}{*}{$\frac{11-5\delta q}{10}$ (1.078)} & \multirow{2}{*}{1.090} & $-2$ & 1 & (10) & $\frac{3-\delta q}{3}$ & \tabularnewline
\cline{13-17} 
 &  &  &  &  &  &  &  &  &  &  &  & $-2$ & 1 & (02) & $\frac{21-5\delta q}{15}$ & \tabularnewline
\hline 
\multirow{2}{*}{+1} & \multirow{2}{*}{$\frac{3}{2}$} & \multirow{2}{*}{(02)} & \multirow{2}{*}{24} & \multirow{2}{*}{$\frac{3}{2}$} & \multirow{2}{*}{+1} & \multirow{2}{*}{$\frac{3}{2}$} & \multirow{2}{*}{(11)} & \multirow{2}{*}{32} & \multirow{2}{*}{$\frac{43+5\delta q}{30}$} & \multirow{2}{*}{$\frac{6}{5}$ (1.200)} & \multirow{2}{*}{1.223} & +1 & $\frac{3}{2}$ & (10) & $\frac{33+5\delta q}{30}$ & \tabularnewline
\cline{13-17} 
 &  &  &  &  &  &  &  &  &  &  &  & +1 & $\frac{3}{2}$ & (02) & $\frac{9+\delta q}{6}$ & \tabularnewline
\hline 
\multirow{2}{*}{$-1$} & \multirow{2}{*}{$\frac{3}{2}$} & \multirow{2}{*}{(20)} & \multirow{2}{*}{24} & \multirow{2}{*}{$\frac{3}{2}$} & \multirow{2}{*}{$-1$} & \multirow{2}{*}{$\frac{3}{2}$} & \multirow{2}{*}{(10)} & \multirow{2}{*}{12} & \multirow{2}{*}{$\frac{33-5\delta q}{30}$} & \multirow{2}{*}{$\frac{13-5\delta q}{15}$ (0.852)} & \multirow{2}{*}{0.844} & $-1$ & $\frac{3}{2}$ & (01) & $\frac{33-5\delta q}{30}$ & \tabularnewline
\cline{13-17} 
 &  &  &  &  &  &  &  &  &  &  &  & $-1$ & $\frac{3}{2}$ & (20) & $\frac{9-\delta q}{6}$ & \tabularnewline
\hline 
\multirow{3}{*}{$\pm3$} & \multirow{3}{*}{$\frac{1}{2}$} & \multirow{3}{*}{(11)} & \multirow{3}{*}{16} & \multirow{3}{*}{$\frac{3}{2}$} & \multirow{2}{*}{$\pm3$} & \multirow{2}{*}{$\frac{1}{2}$} & \multirow{2}{*}{(01)} & \multirow{2}{*}{6} & \multirow{2}{*}{$\frac{7\pm3\delta q}{6}$} & \multirow{2}{*}{$\frac{14\pm5\delta q}{15}$ (0.948/0.919)} & \multirow{2}{*}{0.954/0.894} & $\pm3$ & $\frac{1}{2}$ & (00) & $\frac{9\pm5\delta q}{10}$ & \tabularnewline
\cline{13-17} 
 &  &  &  &  &  &  &  &  &  &  &  & $\pm3$ & $\frac{1}{2}$ & (11) & $\frac{3\pm\delta q}{2}$ & \tabularnewline
\cline{6-17} 
 &  &  &  &  & $\pm3$ & $\frac{1}{2}$ & (20) & 12 & $\frac{47\pm15\delta q}{30}$ & $\frac{4\pm\delta q}{3}$ (1.348/1.319) & 1.432/1.311 & $\pm3$ & $\frac{1}{2}$ & (11) & $\frac{3\pm\delta q}{2}$ & \tabularnewline
\hline 
\multirow{2}{*}{$\pm3$} & \multirow{2}{*}{$\frac{3}{2}$} & \multirow{2}{*}{(00)} & \multirow{2}{*}{4} & \multirow{2}{*}{$\frac{3}{2}$} & \multirow{2}{*}{$\pm3$} & \multirow{2}{*}{$\frac{3}{2}$} & \multirow{2}{*}{(01)} & \multirow{2}{*}{12} & \multirow{2}{*}{$\frac{53\pm15\delta q}{30}$} & \multirow{2}{*}{$\frac{23\pm5\delta q}{15}$ (1.548/1.519)} & \multirow{2}{*}{1.599/1.579} & $\pm3$ & $\frac{3}{2}$ & (00) & $\frac{3\pm\delta q}{2}$ & \tabularnewline
\cline{13-17} 
 &  &  &  &  &  &  &  &  &  &  &  & $\pm3$ & $\frac{3}{2}$ & (11) & $\frac{21\pm5\delta q}{10}$ & \tabularnewline
\hline 
\hline
\end{tabular}
\end{table*}

\begin{table*}
\caption{Same fusion procedure as that in Table~\ref{tab:cft2},
but now compared to NRG data computed at $J_{0}=0.6,K_{0}=0.3,I_{0}=-0.13$, where $[0, 0, (01)]$ becomes the ground state (in contrast to $[+1, \frac{1}{2}, (00)]$ used in Table~\ref{fig:flow2}). The excitation energies are defined as, $\delta E^{\prime}=E^{\prime}-E_{\text{min}}^{\prime}$, where $E_{\text{min}}^{\prime}=E^{\prime}([0,0,(01)])=\frac{4}{15}$. The NRG energies have been
shifted and rescaled such that the ground state is zero and the values
of $E_{\text{NRG}}$ and $\delta E^{\prime}$ match for the multiplet
$[0,1,(01)]$. $\delta q$ is then determined by matching
$E_{\text{NRG}}$ and $\delta E^{\prime}$
for the multiplet $[+1,\frac{1}{2},(00)]$,
resulting in 
$\delta q=0.2135$. 
}
\label{tab:cft3}
\renewcommand{\arraystretch}{1.8}

\begin{tabular}{ccccc|cccccc|c|ccccc}
\hline
\hline
\multicolumn{5}{c|}{free fermions} & \multicolumn{6}{c|}{single fusion, with $Q_{\text{imp}}^{\text{eff}}=[0,0,(01)]$} & NRG & \multicolumn{5}{c}{double fusion, with $\bar{Q}_{\text{imp}}^{\text{eff}}=[0,0,(10)]$}\tabularnewline
$q$ & $S$ & $(\lambda_{1}\lambda_{2})$ & $d$ & $E$ & $q^{\prime}$ & $S^{\prime}$ & $(\lambda_{1}^{\prime}\lambda_{2}^{\prime})$ & $d^{\prime}$ & $E^{\prime}$ & $\delta E^{\prime}$ & $E_{\text{NRG}}$ & $q^{\prime\prime}$ & $S^{\prime\prime}$ & $(\lambda_{1}^{\prime\prime}\lambda_{2}^{\prime\prime})$ & $\Delta$ & $\hat{O}$\tabularnewline
\hline 
\hline
\multirow{2}{*}{0} & \multirow{2}{*}{0} & \multirow{2}{*}{(00)} & \multirow{2}{*}{1} & \multirow{2}{*}{0} & \multirow{2}{*}{0} & \multirow{2}{*}{0} & \multirow{2}{*}{(01)} & \multirow{2}{*}{3} & \multirow{2}{*}{$\frac{4}{15}$} & \multirow{2}{*}{0 } & \multirow{2}{*}{0} & 0 & 0 & (00) & 0 & $\mathds{1}$\tabularnewline
\cline{13-17} 
 &  &  &  &  &  &  &  &  &  &  &  & 0 & 0 & (11) & \textcolor{ForestGreen}{$\frac{3}{5}$ ($=\Delta_{\text{orb}}$)} & $\Phi_{\text{orb}}$\tabularnewline
\hline 
\multirow{3}{*}{+1} & \multirow{3}{*}{$\frac{1}{2}$} & \multirow{3}{*}{(10)} & \multirow{3}{*}{6} & \multirow{3}{*}{$\frac{1}{2}$} & +1 & $\frac{1}{2}$ & (00) & 2 & $\frac{7+5\delta q}{30}$ & $\frac{-1+5\delta q}{30}$ (0.002) & 0.002 & +1 & $\frac{1}{2}$ & (10) & $\frac{3+\delta q}{6}$ & \tabularnewline
\cline{6-17} 
 &  &  &  &  & \multirow{2}{*}{+1} & \multirow{2}{*}{$\frac{1}{2}$} & \multirow{2}{*}{(11)} & \multirow{2}{*}{16} & \multirow{2}{*}{$\frac{5+\delta q}{6}$} & \multirow{2}{*}{$\frac{17+5\delta q}{30}$(0.602)} & \multirow{2}{*}{0.613} & +1 & $\frac{1}{2}$ & (10) & $\frac{3+\delta q}{6}$ & \tabularnewline
\cline{13-17} 
 &  &  &  &  &  &  &  &  &  &  &  & +1 & $\frac{1}{2}$ & (02) & $\frac{27+5\delta q}{30}$ & \tabularnewline
\hline 
\multirow{3}{*}{$-1$} & \multirow{3}{*}{$\frac{1}{2}$} & \multirow{3}{*}{(01)} & \multirow{3}{*}{6} & \multirow{3}{*}{$\frac{1}{2}$} & \multirow{2}{*}{$-1$} & \multirow{2}{*}{$\frac{1}{2}$} & \multirow{2}{*}{(10)} & \multirow{2}{*}{6} & \multirow{2}{*}{$\frac{3-\delta q}{6}$} & \multirow{2}{*}{$\frac{7-5\delta q}{30}$ (0.198)} & \multirow{2}{*}{0.196} & $-1$ & $\frac{1}{2}$ & (01) & $\frac{3-\delta q}{6}$ & \tabularnewline
\cline{13-17} 
 &  &  &  &  &  &  &  &  &  &  &  & $-1$ & $\frac{1}{2}$ & (20) & $\frac{27-5\delta q}{30}$ & \tabularnewline
\cline{6-17} 
 &  &  &  &  & $-1$ & $\frac{1}{2}$ & (02) & 12 & $\frac{27-5\delta q}{30}$ & $\frac{19-5\delta q}{30}$ (0.598) & 0.605 & $-1$ & $\frac{1}{2}$ & (01) & $\frac{3-\delta q}{6}$ & \tabularnewline
\hline 
\multirow{3}{*}{0} & \multirow{3}{*}{1} & \multirow{3}{*}{(11)} & \multirow{3}{*}{24} & \multirow{3}{*}{1} & \multirow{2}{*}{0} & \multirow{2}{*}{1} & \multirow{2}{*}{(01)} & \multirow{2}{*}{9} & \multirow{2}{*}{$\frac{2}{3}$} & \multirow{2}{*}{$\frac{2}{5}$ (0.400)} & \multirow{2}{*}{0.400} & 0 & 1 & (00) & \textcolor{red}{$\frac{2}{5}$ ($=\Delta_{\text{sp}}$)} & $\Phi_{\text{sp}}$\tabularnewline
\cline{13-17} 
 &  &  &  &  &  &  &  &  &  &  &  & 0 & 1 & (11) & \textcolor{blue}{1 ($=\Delta_{\text{sp-orb}}$)} & $\Phi_{\text{sp-orb}}$ \tabularnewline
\cline{6-17} 
 &  &  &  &  & 0 & 1 & (20) & 18 & $\frac{16}{15}$ & $\frac{4}{5}$ (0.800) & 0.812 & 0 & 1 & (11) & 1 & \tabularnewline
\hline 
\multirow{2}{*}{+2} & \multirow{2}{*}{0} & \multirow{2}{*}{(20)} & \multirow{2}{*}{6} & \multirow{2}{*}{1} & \multirow{2}{*}{+2} & \multirow{2}{*}{0} & \multirow{2}{*}{(10)} & \multirow{2}{*}{3} & \multirow{2}{*}{$\frac{9+5\delta q}{15}$} & \multirow{2}{*}{$\frac{1+\delta q}{3}$ (0.405)} & \multirow{2}{*}{0.407} & +2 & 0 & (01) & $\frac{9+5\delta q}{15}$ & \tabularnewline
\cline{13-17} 
 &  &  &  &  &  &  &  &  &  &  &  & +2 & 0 & (20) & $\frac{3+\delta q}{3}$ & \tabularnewline
\hline 
\multirow{2}{*}{$-2$} & \multirow{2}{*}{0} & \multirow{2}{*}{(02)} & \multirow{2}{*}{6} & \multirow{2}{*}{1} & \multirow{2}{*}{$-2$} & \multirow{2}{*}{0} & \multirow{2}{*}{(11)} & \multirow{2}{*}{8} & \multirow{2}{*}{$\frac{14-5\delta q}{15}$} & \multirow{2}{*}{$\frac{2-\delta q}{3}$ (0.596)} & \multirow{2}{*}{0.595} & $-2$ & 0 & (10) & $\frac{9-5\delta q}{15}$ & \tabularnewline
\cline{13-17} 
 &  &  &  &  &  &  &  &  &  &  &  & $-2$ & 0 & (02) & $\frac{3-\delta q}{3}$ & \tabularnewline
\hline 
\multirow{3}{*}{+2} & \multirow{3}{*}{1} & \multirow{3}{*}{(01)} & \multirow{3}{*}{9} & \multirow{3}{*}{1} & \multirow{2}{*}{+2} & \multirow{2}{*}{1} & \multirow{2}{*}{(10)} & \multirow{2}{*}{9} & \multirow{2}{*}{$\frac{3+\delta q}{3}$} & \multirow{2}{*}{$\frac{11+5\delta q}{15}$ (0.805)} & \multirow{2}{*}{0.859} & +2 & 1 & (01) & $\frac{3+\delta q}{3}$ & \tabularnewline
\cline{13-17} 
 &  &  &  &  &  &  &  &  &  &  &  & +2 & 1 & (20) & $\frac{21+5\delta q}{15}$ & \tabularnewline
\cline{6-17} 
 &  &  &  &  & +2 & 1 & (02) & 18 & $\frac{21+5\delta q}{15}$ & $\frac{17+5\delta q}{15}$ (1.205) & 1.234 & +2 & 1 & (01) & $\frac{3+\delta q}{3}$ & \tabularnewline
\hline 
\multirow{3}{*}{$-2$} & \multirow{3}{*}{1} & \multirow{3}{*}{(10)} & \multirow{3}{*}{9} & \multirow{3}{*}{1} & $-2$ & 1 & (00) & 3 & $\frac{11-5\delta q}{15}$ & $\frac{7-5\delta q}{15}$ (0.396) & 0.391 & $-2$ & 1 & (10) & $\frac{3-\delta q}{3}$ & \tabularnewline
\cline{6-17} 
 &  &  &  &  & \multirow{2}{*}{$-2$} & \multirow{2}{*}{1} & \multirow{2}{*}{(11)} & \multirow{2}{*}{24} & \multirow{2}{*}{$\frac{4-\delta q}{3}$} & \multirow{2}{*}{$\frac{16-5\delta q}{15}$ (0.996)} & \multirow{2}{*}{1.015} & $-2$ & 1 & (10) & $\frac{3-\delta q}{3}$ & \tabularnewline
\cline{13-17} 
 &  &  &  &  &  &  &  &  &  &  &  & $-2$ & 1 & (02) & $\frac{21-5\delta q}{15}$ & \tabularnewline
\hline 
\multirow{2}{*}{+1} & \multirow{2}{*}{$\frac{3}{2}$} & \multirow{2}{*}{(02)} & \multirow{2}{*}{24} & \multirow{2}{*}{$\frac{3}{2}$} & \multirow{2}{*}{+1} & \multirow{2}{*}{$\frac{3}{2}$} & \multirow{2}{*}{(11)} & \multirow{2}{*}{32} & \multirow{2}{*}{$\frac{43+5\delta q}{30}$} & \multirow{2}{*}{$\frac{7+\delta q}{6}$ (1.202)} & \multirow{2}{*}{1.249} & +1 & $\frac{3}{2}$ & (10) & $\frac{33+5\delta q}{30}$ & \tabularnewline
\cline{13-17} 
 &  &  &  &  &  &  &  &  &  &  &  & +1 & $\frac{3}{2}$ & (02) & $\frac{9+\delta q}{6}$ & \tabularnewline
\hline 
\multirow{2}{*}{$-1$} & \multirow{2}{*}{$\frac{3}{2}$} & \multirow{2}{*}{(20)} & \multirow{2}{*}{24} & \multirow{2}{*}{$\frac{3}{2}$} & \multirow{2}{*}{$-1$} & \multirow{2}{*}{$\frac{3}{2}$} & \multirow{2}{*}{(10)} & \multirow{2}{*}{12} & \multirow{2}{*}{$\frac{33-5\delta q}{30}$} & \multirow{2}{*}{$\frac{5-\delta q}{6}$ (0.798)} & \multirow{2}{*}{0.799} & $-1$ & $\frac{3}{2}$ & (01) & $\frac{33-5\delta q}{30}$ & \tabularnewline
\cline{13-17} 
 &  &  &  &  &  &  &  &  &  &  &  & $-1$ & $\frac{3}{2}$ & (20) & $\frac{9-\delta q}{6}$ & \tabularnewline
\hline 
\multirow{3}{*}{$\pm3$} & \multirow{3}{*}{$\frac{1}{2}$} & \multirow{3}{*}{(11)} & \multirow{3}{*}{16} & \multirow{3}{*}{$\frac{3}{2}$} & \multirow{2}{*}{$\pm3$} & \multirow{2}{*}{$\frac{1}{2}$} & \multirow{2}{*}{(01)} & \multirow{2}{*}{6} & \multirow{2}{*}{$\frac{7\pm3\delta q}{6}$} & \multirow{2}{*}{$\frac{9\pm5\delta q}{10}$ (1.007/0.793)} & \multirow{2}{*}{1.033/0.791} & $\pm3$ & $\frac{1}{2}$ & (00) & $\frac{9\pm5\delta q}{10}$ & \tabularnewline
\cline{13-17} 
 &  &  &  &  &  &  &  &  &  &  &  & $\pm3$ & $\frac{1}{2}$ & (11) & $\frac{3\pm\delta q}{2}$ & \tabularnewline
\cline{6-17} 
 &  &  &  &  & $\pm3$ & $\frac{1}{2}$ & (20) & 12 & $\frac{47\pm15\delta q}{30}$ & $\frac{13\pm5\delta q}{10}$ (1.407/1.193) & 1.527/1.209 & $\pm3$ & $\frac{1}{2}$ & (11) & $\frac{3\pm\delta q}{2}$ & \tabularnewline
\hline 
\multirow{2}{*}{$\pm3$} & \multirow{2}{*}{$\frac{3}{2}$} & \multirow{2}{*}{(00)} & \multirow{2}{*}{4} & \multirow{2}{*}{$\frac{3}{2}$} & \multirow{2}{*}{$\pm3$} & \multirow{2}{*}{$\frac{3}{2}$} & \multirow{2}{*}{(01)} & \multirow{2}{*}{12} & \multirow{2}{*}{$\frac{53\pm15\delta q}{30}$} & \multirow{2}{*}{$\frac{3\pm\delta q}{2}$ (1.607/1.393)} & \multirow{2}{*}{1.695/1.474} & $\pm3$ & $\frac{3}{2}$ & (00) & $\frac{3\pm\delta q}{2}$ & \tabularnewline
\cline{13-17} 
 &  &  &  &  &  &  &  &  &  &  &  & $\pm3$ & $\frac{3}{2}$ & (11) & $\frac{21\pm5\delta q}{10}$ & \tabularnewline
\hline 
\hline
\end{tabular}
\end{table*}

\begin{table*}
\caption{Demonstration that the $\it{NFL}^{\prime}$ fixed point spectrum is the union of the FL and NFL fixed point spectra near the phase transition. Second column: $E_{{\it NFL}^{\prime}}$ of the intermediate-energy regime ${\it NFL}^{\prime}$ at $I_{0}=-0.00609896199692$ [Fig.~3(b) and Fig.~\ref{fig:flow2}(b)]. Third column: the low-energy FL spectrum with ground state $[-2,0,(00)]$
just before the phase transition at $I_{0}=-0.006098$ [Fig.~3(a) and Fig.~\ref{fig:flow2}(a)]. Fourth column: the low-energy NFL spectrum with ground state $[+1,\tfrac{1}{2},(00)]$
just after the phase transition at $I_{0}=-0.006099$ [Fig.~3(c) and Fig.~\ref{fig:flow2}(c)].
All entries that relate or can be linked to FL are marked in red.
The FL excitations can be obtained by adding electrons with energy $\epsilon_{e}=0.275511$
or holes with energy $\epsilon_{p}=1.553042$ on top of the ground
state with energy $E_g$, as indicated in the column for $E_\mathrm{FL}$.}
\label{tab:nfl_prime}
\renewcommand{\arraystretch}{2.0}
\begin{tabular}{cccc|c|c|c}
\hline 
\hline
\multicolumn{4}{c|}{Multiplets} & $E_{\text{{\it {NFL}}}^{\prime}}$  & \textcolor{red}{$E_{\text{FL}}$ } & $E_{\text{NFL}}$ \tabularnewline
$q$ & $S$ & ($\lambda_{1}\lambda_{2}$) & $d$ & @$I_{0}=-0.00609896199692$ & \textcolor{red}{@$I_{0}=-0.006098$} & @$I_{0}=-0.006099$\tabularnewline
\hline 
\hline
+1 & $\frac{1}{2}$ & (00) & 2 & $-0.717930$ &  & $-0.729529$ ($=E_g$)
\tabularnewline
\hline 
0 & 0 & (01) & 3 & $-0.668562$ &  & $-0.680530$\tabularnewline
\hline 
\textcolor{red}{$-2$} & \textcolor{red}{0} & \textcolor{red}{(00)} & \textcolor{red}{1} & \textcolor{red}{$-0.550738$} & \textcolor{red}{$-0.551028$ ($=E_{g}$)} & \tabularnewline
\hline 
\textcolor{red}{$-1$} & \textcolor{red}{$\frac{1}{2}$} & \textcolor{red}{(10)} & \textcolor{red}{6} & \textcolor{red}{$-0.282883$} & \textcolor{red}{$-0.275517$ ($\simeq E_{g}+\epsilon_{e}$)} & \tabularnewline
\hline 
\textcolor{black}{$-1$} & \textcolor{black}{$\frac{1}{2}$} & \textcolor{black}{(10)} & \textcolor{black}{6} & \textcolor{black}{$-0.275324$} &  & $-0.295054$\tabularnewline
\hline 
+2 & 0 & (10) & 3 & $-0.0880684$ &  & $-0.100011$\tabularnewline
\hline 
\textcolor{red}{0} & \textcolor{red}{1} & \textcolor{red}{(01)} & \textcolor{red}{9} & \textcolor{red}{0} & \textcolor{red}{0 ($\simeq E_{g}+2\epsilon_{e}$)} & \tabularnewline
\hline 
\textcolor{red}{0} & \textcolor{red}{0} & \textcolor{red}{(20)} & \textcolor{red}{6} & \textcolor{red}{0.000037} & \textcolor{red}{0 ($\simeq E_{g}+2\epsilon_{e}$)} & \tabularnewline
\hline 
0 & 1 & (01) & 9 & 0.011703 &  & 0\tabularnewline
\hline 
$-2$ & 1 & (00) & 3 & 0.102149 &  & 0.089692\tabularnewline
\hline 
\textcolor{red}{+1} & \textcolor{red}{$\frac{3}{2}$} & \textcolor{red}{(00)} & \textcolor{red}{4} & \textcolor{red}{0.275407} & \textcolor{red}{0.275518 ($\simeq E_{g}+3\epsilon_{e}$)} & \tabularnewline
\hline 
\textcolor{red}{+1} & \textcolor{red}{$\frac{1}{2}$} & \textcolor{red}{(11)} & \textcolor{red}{16} & \textcolor{red}{0.275410} & \textcolor{red}{0.275518 ($\simeq E_{g}+3\epsilon_{e}$)} & \tabularnewline
\hline 
+1 & $\frac{1}{2}$ & (11) & 16 & 0.313029 &  & 0.300831\tabularnewline
\hline 
$-1$ & $\frac{1}{2}$ & (02) & 12 & 0.416507 &  & 0.404921\tabularnewline
\hline 
$-2$ & 0 & (11) & 8 & 0.451411 &  & 0.439442\tabularnewline
\hline 
\textcolor{red}{+2} & \textcolor{red}{1} & \textcolor{red}{(10)} & \textcolor{red}{9} & \textcolor{red}{0.550783} & \textcolor{red}{0.551033 ($\simeq E_{g}+4\epsilon_{e}$)} & \tabularnewline
\hline 
\textcolor{red}{+2} & \textcolor{red}{0} & \textcolor{red}{(02)} & \textcolor{red}{6} & \textcolor{red}{0.550791} & \textcolor{red}{0.551033 ($\simeq E_{g}+4\epsilon_{e}$)} & \tabularnewline
\hline 
+2 & 1 & (10) & 9 & 0.666084 &  & 0.654338\tabularnewline
\hline 
0 & 1 & (20) & 18 & 0.703088 &  & 0.690623\tabularnewline
\hline 
$-1$ & $\frac{3}{2}$ & (10) & 12 & 0.741511 &  & 0.729522\tabularnewline
\hline 
\textcolor{red}{+3} & \textcolor{red}{$\frac{1}{2}$} & \textcolor{red}{(01)} & \textcolor{red}{6} &
\textcolor{red}{0.826154} 
& \textcolor{red}{0.826543 ($\simeq E_{g}+5\epsilon_{e}$)} & \tabularnewline
\hline 
$-3$ & $\frac{1}{2}$ & (01) & 6 & 0.835829 &  & 0.823640\tabularnewline
\hline 
+3 & $\frac{1}{2}$ & (01) & 6 & 0.911424 &  & 0.899222\tabularnewline
\hline 
\textcolor{red}{$-3$} & \textcolor{red}{$\frac{1}{2}$} & \textcolor{red}{(01)} & \textcolor{red}{6} & \textcolor{red}{1.002497} & \textcolor{red}{1.002014 ($\simeq E_{g}+\epsilon_{p}$)} & \tabularnewline
\hline 
\hline
\end{tabular}
\end{table*}

\section{Interpretation of $\it{NFL}^\prime$ regime}
The $\it{NFL}^\prime$ regime appears as an intermediate fixed point in the NRG flow diagram [see Fig.~3(b) in the main text and Fig.~\ref{fig:flow2}(b,e) in this supplemental material], when the system is close to the phase boundary between the FL and NFL phases.
The $\it{NFL}^\prime$ fixed-point spectrum is a ``superposition" of the FL and NFL fixed-point spectra.
To be more precise, the set of lowest-lying energy levels in the $\it{NFL}^\prime$ regime [Figs.~\ref{fig:flow2}(b,e)] is the union of the sets of the levels in the FL regime [Figs.~\ref{fig:flow2}(a,d)] and in the NFL regime [Figs.~\ref{fig:flow2}(c,f)] at iterations of the same parity (even or odd length). Table~\ref{tab:nfl_prime} shows this by listing the energy spectrum of $\it{NFL}^{\prime}$ for a Wilson chain with odd length at $J_0=0.3, I_{0}=-0.00609896199692$ [Figs.~\ref{fig:flow2}(b)] and comparing it with the FL spectrum just before the phase transition at $I_{0}=-0.006098$ [Figs.~\ref{fig:flow2}(a)] and the NFL spectrum just after the phase transition at $I_{0}=-0.006099$ [Figs.~\ref{fig:flow2}(c)].
When the system is slightly away from the phase boundary, there is a reduction from $\it{NFL}^\prime$ to FL or NFL.
During this `crossover', a set of levels for either FL or NFL regime remains in the low-energy part of the NRG flow diagram, while the other set abruptly drifts towards higher energies due to the intrinsic exponential rescaling of NRG flow diagrams.
Such an abrupt drift of a subset of levels is in stark contrast to the typically encountered smooth tangled fixed-point crossovers
in impurity models. 

The superposition nature of the $\it{NFL}^\prime$ fixed-point spectrum and the abrupt crossover indicate that the phase transition between the FL and NFL phase follows the ``level-crossing'' scenario.
Within this scenario, 
the low-energy sector of the spectrum is the union of orthogonal subspaces.
One subspace consists of the 
states $| E_i^\mathrm{FL} \rangle$
at the FL fixed point, the other 
of the 
states $| E_i^\mathrm{NFL} \rangle$
at the NFL fixed point, with their respective ground states at $i=0$.
At energy scales $\omega, T$ larger than the difference of the ground-state energies $| E_0^\mathrm{FL} - E_0^\mathrm{NFL} |$, both subspaces contribute to thermodynamic and dynamical properties, such as the impurity contribution to entropy $S_\mathrm{imp}$ and dynamical impurity susceptibility $\chi^\mathrm{imp}$.
This corresponds to the $\it{NFL}^\prime$ regime appearing at earlier iterations in the NRG flow diagrams.
As one proceeds with the NRG steps, 
the energy scale becomes smaller and eventually reaches 
$| E_0^\mathrm{FL} - E_0^\mathrm{NFL} |$.
At this point, the higher lying fixed point starts to disappear
from the low-energy physics, as clearly visible in the NRG flow diagram in Figs.~\ref{fig:flow2}(b,e).
As an aside, we note that a similar 
level-crossing competition between two subspaces has been
used to induce a two-stage Kondo effect in driven quantum dot systems~\cite{sbierski:2013,lee:2019}.

With $S_\mathrm{imp} \equiv \ln g$, where $g^\mathrm{FL}{=}1$
and \mbox{$g^\mathrm{NFL}{=}\tfrac{1}{2}(1+\sqrt{5})$},
the level-crossing scenario leads to an additive behavior
of $g$, i.e., 
\begin{eqnarray}
S_\mathrm{imp}^{\mathrm{NFL}^\prime} &=& \ln (
g^\mathrm{FL} + g^\mathrm{NFL})
\text{ ,}\label{eq:gadd}
\end{eqnarray}
%
which follows from elementary considerations 
based on the definition of the partition function.
To derive this relation, we start from 
the partition function of the whole system (impurity plus bath).
In the $\it{NFL}^\prime$ regime, 
\begin{equation}
\begin{aligned}
Z^{\mathrm{NFL}^\prime} &= Z^{\mathrm{FL}} + Z^{\mathrm{NFL}},
\end{aligned}
\label{eq:Z_NFLprime}
\end{equation}
with $Z^\alpha {\equiv} \sum_i \exp (- E_i^{\alpha}/T)$
at temperature $T$. This 
is the {\it sum} 
of partition functions for the NFL and FL regimes,
since the FL and NFL subspaces are orthogonal to each other.
Now the entropy is related to the partition function
via the free energy $F = -T \ln Z$, having $S=-\tfrac{\partial}{\partial T} F$.
The impurity contribution to the entropy is defined as
the difference between the entropy of the whole system
(impurity plus bath) and the entropy of the bath only,
\begin{subequations}
\begin{eqnarray}
S_\mathrm{imp} &\equiv& 
S_\mathrm{tot} - S_\mathrm{bath} =
\tfrac{\partial}{\partial T} {\bigl[ T \bigl( \ln Z_\mathrm{tot} -
\ln Z_\mathrm{bath}) \bigr) \bigr]
}
\notag \\
&=& \ln \tfrac{Z_\mathrm{tot}}{Z_\mathrm{bath}} 
 +
\underbrace{\tfrac{1}{T} \bigl(  
U_\mathrm{tot} - U_\mathrm{bath}
\bigr)}_{\simeq \ \mathrm{const} \ \cong \ \ln a }
\label{eq:S_imp:2} \\
&\simeq& \ln \bigl[ \tfrac{a\,Z_\mathrm{tot}}{Z_\mathrm{bath}} \bigr]
\text{ .} \label{eq:S_imp:3}
\end{eqnarray}
\end{subequations}
where 
$U {=} T^2 \tfrac{\partial}{\partial T} \ln Z  {\equiv} \langle E \rangle$
is the internal 
energy.
The entropy is independent of the arbitrary choice of
energy reference, as also apparent from the first line.
Therefore both $Z_\mathrm{tot}$ and
$Z_\mathrm{bath}$ can be computed relative to their
respective ground state energies.
Generally for gapless spectra then,
the expectation values take $U_\mathrm{tot} \sim U_\mathrm{bath} \sim T$.
Moreover, in the level-crossing scenario in the regime
$| E_0^\mathrm{FL} - E_0^\mathrm{NFL} | \ll T$ (relative
to the same energy reference), the two
phases FL and NFL coexist, such that they also show similar
energetics, i.e., $| U_\mathrm{FL} - U_\mathrm{NFL} | \ll T$.
Therefore with $U_\mathrm{tot} = b U_\mathrm{FL} + (1-b)
U_\mathrm{NFL}\simeq \mathrm{const}$ for arbitrary $b\in[0,1]$,
the last term in Eq.\,\eqref{eq:S_imp:2} resembles a constant,
irrespective of having {\it NFL}$'$, NFL, or FL.
Eq.~\eqref{eq:S_imp:3} implies that $g^x=e^{S^x_{\text{imp}}}=a\frac{Z^x_{\text{tot}}}{Z_{\text{bath}}}$ for each of $x=$ FL, NFL and $\it{NFL}^{\prime}$. Eq.~\eqref{eq:gadd} for $S_{\text{imp}}^{\it{NFL}^{\prime}}$ then directly follows from Eq.~\eqref{eq:Z_NFLprime}.

%

The level-crossing scenario also explains
the same power laws of impurity susceptibilities $\chi^\mathrm{imp}$ in the NFL and $\it{NFL}^\prime$ regimes
as well as the kinks of the susceptibilities at the crossover from the $\it{NFL}^\prime$ regime to the NFL regime [Fig.~\ref{fig:flow2}(i)].
In the $\it{NFL}^\prime$ regime, both FL and NFL subspaces contribute to the susceptibility independently,
\begin{equation}
\chi^\mathrm{imp} \simeq (\chi^\mathrm{FL} + \chi^\mathrm{NFL})/2.
\end{equation}
For spin and orbital susceptibilities, the power exponents for $\chi^\mathrm{NFL}$ are smaller than 1, which is the power of $\chi^\mathrm{FL}$. Since the $\it{NFL}^\prime$ regime is already at low energies, the power law of $\chi^\mathrm{FL} + \chi^\mathrm{NFL}$ is dominated by that of $\chi^\mathrm{NFL}$ ($\omega^{-1/5}, \omega^{1/5} \gg \omega^{1}$ when $\omega<1$). At the crossover from the $\it{NFL}^\prime$ regime to the NFL regime,
the contribution of $\chi^\mathrm{FL}$ disappears, which results in the kinks.

\section{Impurity spectral functions}

The impurity spectral functions are calculated via the $\mathcal{T}$-matrix \cite{Costi00,koller:2005},  resulting in the local correlation function,
\begin{equation}
    A_t(\omega)=-\tfrac{1}{\pi} \, \text{Im}\Bigl(
\pi^2\rho_0 \langle O_{m\sigma}||O_{m\sigma}^{\dagger}\rangle_{\omega}\Bigr),
\end{equation}
in the composite operator
$O_{m\sigma}=[\psi_{m\sigma}, H_{\text{int}}]$, 
where $\rho_0=\frac{1}{2D}$ is the bare density of states with $D:=1$ the half-band width of the bath. The normalization ensures that
$A_t \in [0,1]$ with $A_t=1$ implying perfect transmission.

\begin{figure}
\includegraphics[width=0.48\textwidth]{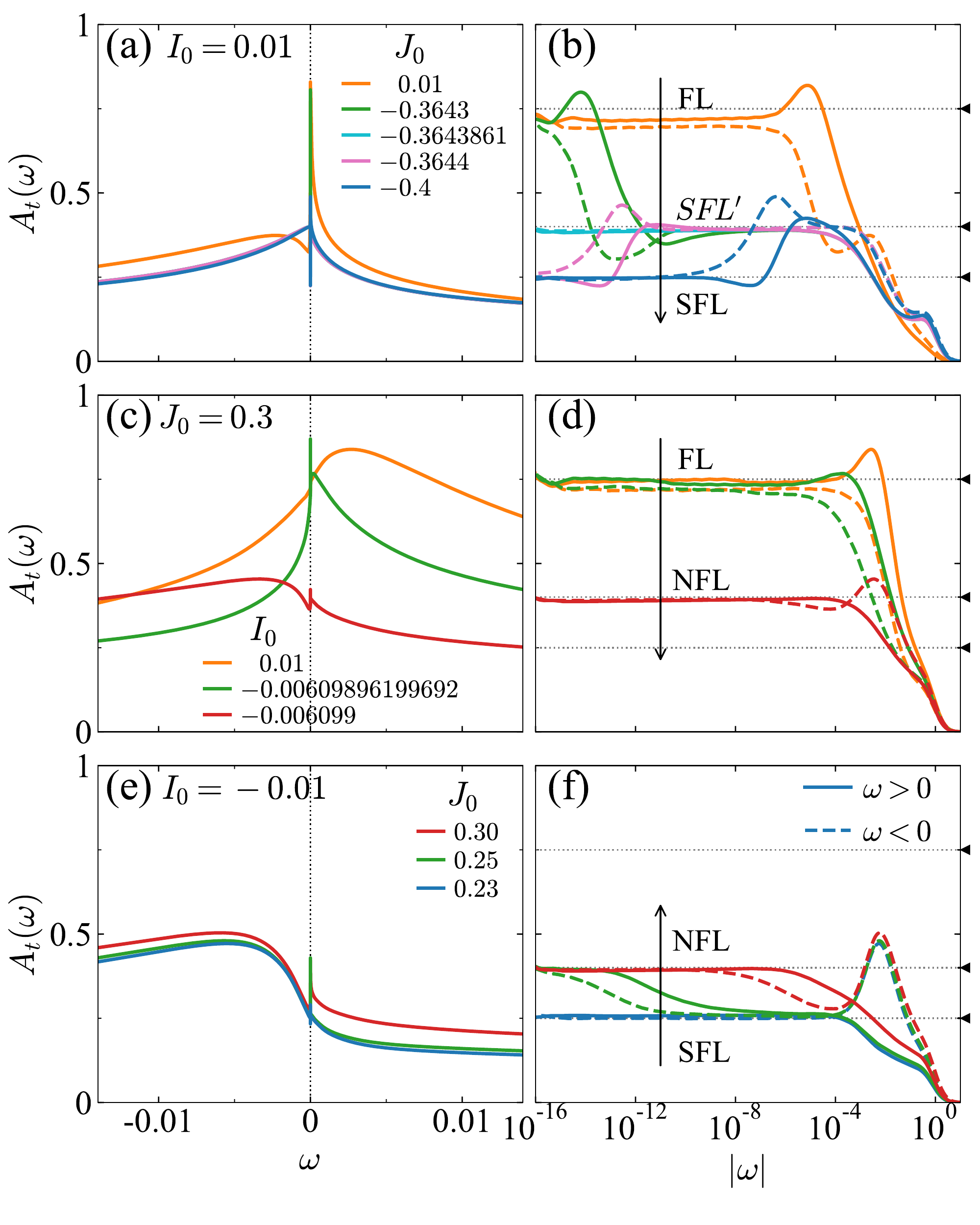}
\caption{Impurity spectral function $A_t(\omega)$. 
         (a,b) transition from FL to SFL,
         (c,d) transition from FL to NFL,
         (e,f) transition from SFL to NFL.
         Lin-lin (Lin-log) plots in the left (right) column.
         In the right column, solid (dashed) lines are for $\omega>0$ ($\omega<0$).
         The grey dotted lines and the black arrows mark the spectral weight of 0.75, 0.4 and 0.25, respectively.}
\label{fig:rho}
\end{figure}

All of the spectral functions scale an asymmetric shape due to the absence of particle-hole symmetry.
In the transition from FL to SFL [Fig.~\ref{fig:rho}(a,b)], the spectral weight is close to the value 0.75 in the FL regime,
while it decreases to about 0.4 in the crossover regime $SFL^\prime$ and further decreases to about 0.25 after entering the SFL regime.
The spectral weight shows a jump from FL to NFL [Fig.~\ref{fig:rho}(c,d)]. In the NFL regime, it is close to the value 0.4, which shows similarity with the crossover regime $SFL^{\prime}$. The evolution of the spectral weight in the transition from SFL to NFL is smooth. In Figs.~\ref{fig:rho}(e,f), we see the crossover of the spectral weight from 0.25 at high energy to 0.4 at low energy for $J_0=0.25$.

\section{The coefficient of impurity specific heat $\gamma=C_{\text{imp}}/T$}
\begin{figure*}
    \centering
    \includegraphics[width=0.95\textwidth]{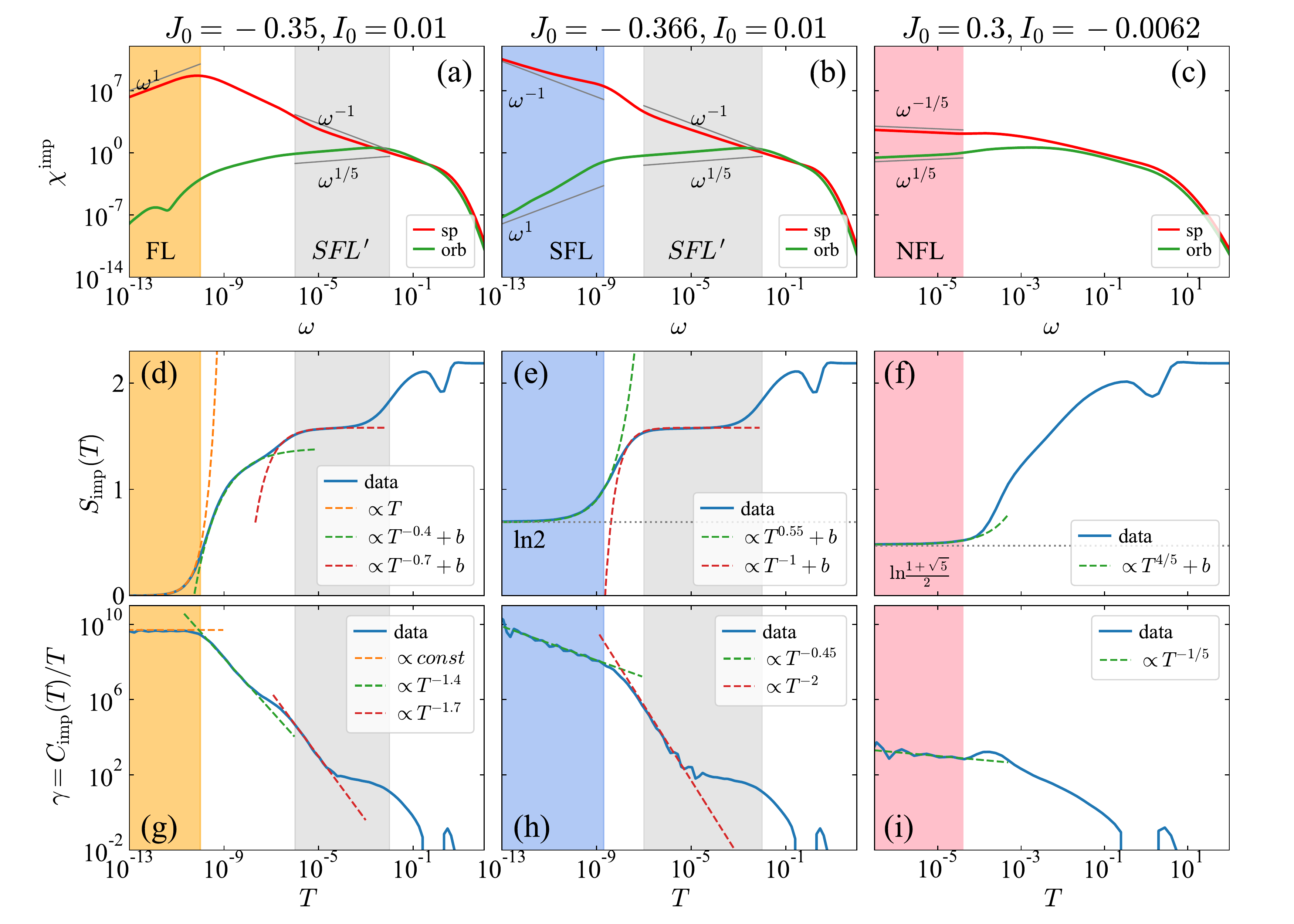}
    \caption{(a-c) Log-log plots of the impurity dynamical susceptibilities of spin and orbital $\chi^{\text{imp}}_{\text{sp,orb}}(\omega)$ as a function of $\omega$ at temperature $T=10^{-16}$. (d-f) Lin-log plots of the impurity contribution to the entropy $S_{\text{imp}}$ as a function of $T$. (g-i) Log-log plots of the coefficient of the impurity contribution to the specific heat $\gamma(T)=C_{\text{imp}}(T)/T$ as a function of $T$. The low-energy fixed points are FL (left), SFL (middle) and NFL (right).}
    \label{fig:gammaT}
\end{figure*}

Figs.~\ref{fig:gammaT}(g-i) show the coefficient of the specific heat,
\begin{equation}
    \gamma(T)=\frac{C_{\text{imp}}(T)}{T}=\frac{\partial{S_{\text{imp}}(T)}}{\partial{T}},
\end{equation}
as a function of temperature $T$ for the FL, SFL and NFL fixed points. The corresponding impurity entropy $S_{\text{imp}}$ [Figs.~\ref{fig:gammaT}(d-f)] and the impurity dynamical susceptibilities of spin and orbital [Figs.~\ref{fig:gammaT}(a-c)] are also plotted for comparison.  $\gamma(T)$ takes a constant value for FL [Fig.~\ref{fig:gammaT}(g)], as expected, while it follows an approximate power law behavior to diverge for SFL [Fig.~\ref{fig:gammaT}(h)] and NFL [Fig.~\ref{fig:gammaT}(i)]. For NFL, the power-law exponent is found to be $-1/5$, which can be obtained by the CFT argument for a SU(2)$_3$ Kondo model, $\gamma(T)\propto T^{\frac{2-3}{2+3}}$, see~\cite{parcollet:1998}.

\section{The impurity static susceptibilities $\chi^{\text{static}}(T)$}
\begin{figure*}
    \centering
    \includegraphics[width=0.95\textwidth]{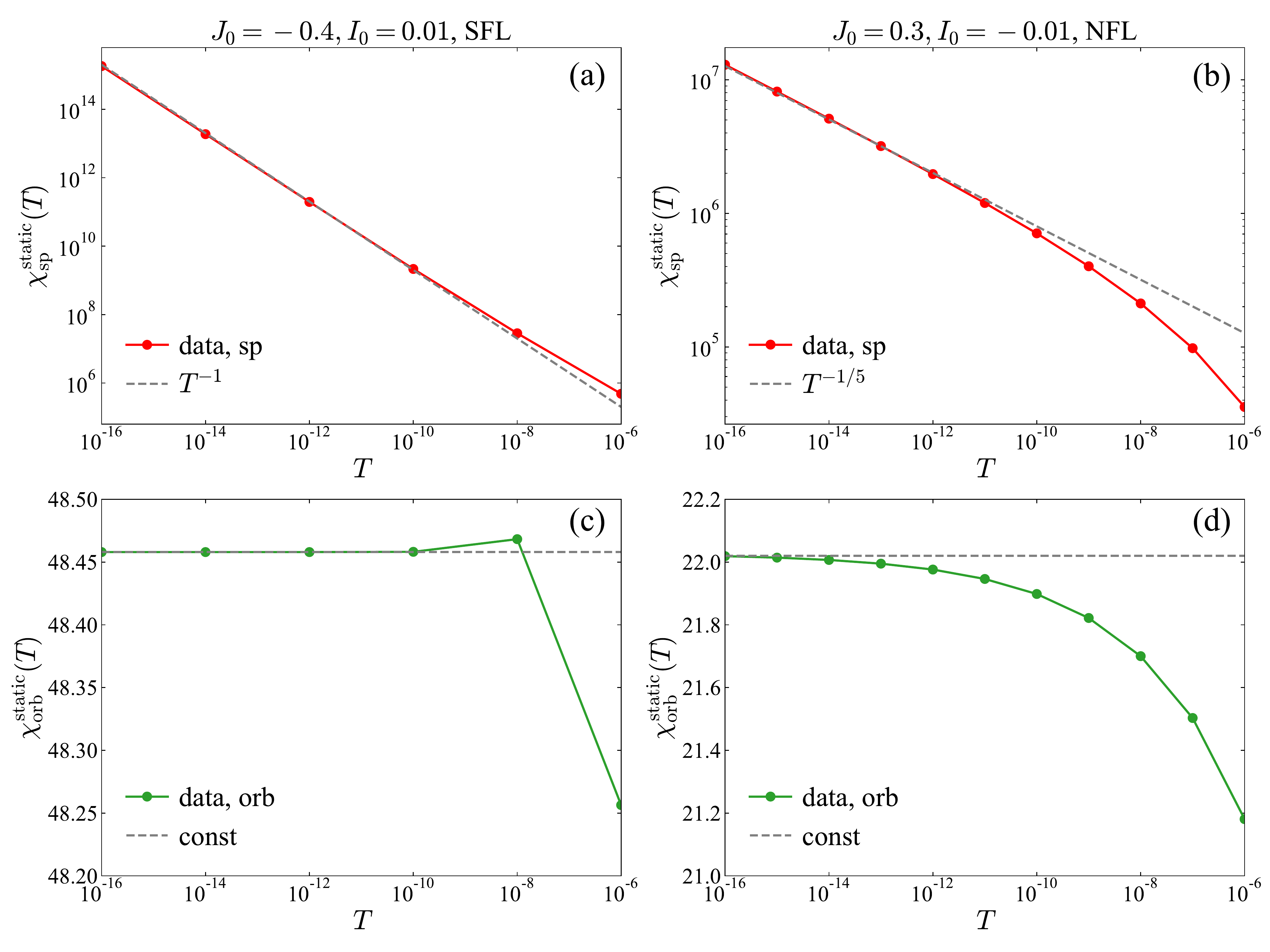}
    \caption{(a-b) Log-log plots of the impurity static susceptibilities of spin $\chi^{\text{static}}_{\text{sp}}(T)$ as functions of temperature $T$. (c-d) Lin-log plots of the impurity static susceptibilities of orbital $\chi^{\text{static}}_{\text{orb}}(T)$ as functions of temperature $T$. (a,c) for SFL and (b,d) for NFL.}
    \label{fig:static}
\end{figure*}

Fig.~\ref{fig:static} shows the impurity static susceptibilities $\chi^{\text{static}}_{\text{sp,orb}}(T)$ as functions of temperature $T$ for SFL (a,c) and NFL (b,d) phases. 
For SFL, $\chi^{\text{static}}(T)$ follow $1/T$ behavior for spin [Fig.~\ref{fig:static}(a)] and constant for orbital [Fig.~\ref{fig:static}(c)] at low temperature, as expected for a decoupled impurity spin moment-$\frac{1}{2}$ and fully screened orbitals. For NFL, it follows $T^{-1/5}$ behavior for spin [Fig.~\ref{fig:static}(b)] at low temperature, which can be obtained by the CFT arguments for a SU(2)$_3$ Kondo model, $\chi^{\text{static}}_{\text{sp}}\propto T^{\frac{2-3}{2+3}}$, see~\cite{parcollet:1998}. While, for orbital, it follows a complex behavior to approach a constant very slowly at low temperature. 

\section{Energy difference of multiplets $[+1, \frac{1}{2}, (00)]$ and $[+1, \frac{3}{2}, (00)]$ along the phase boundary}
\begin{figure}
    \centering
    \includegraphics[width=0.48\textwidth]{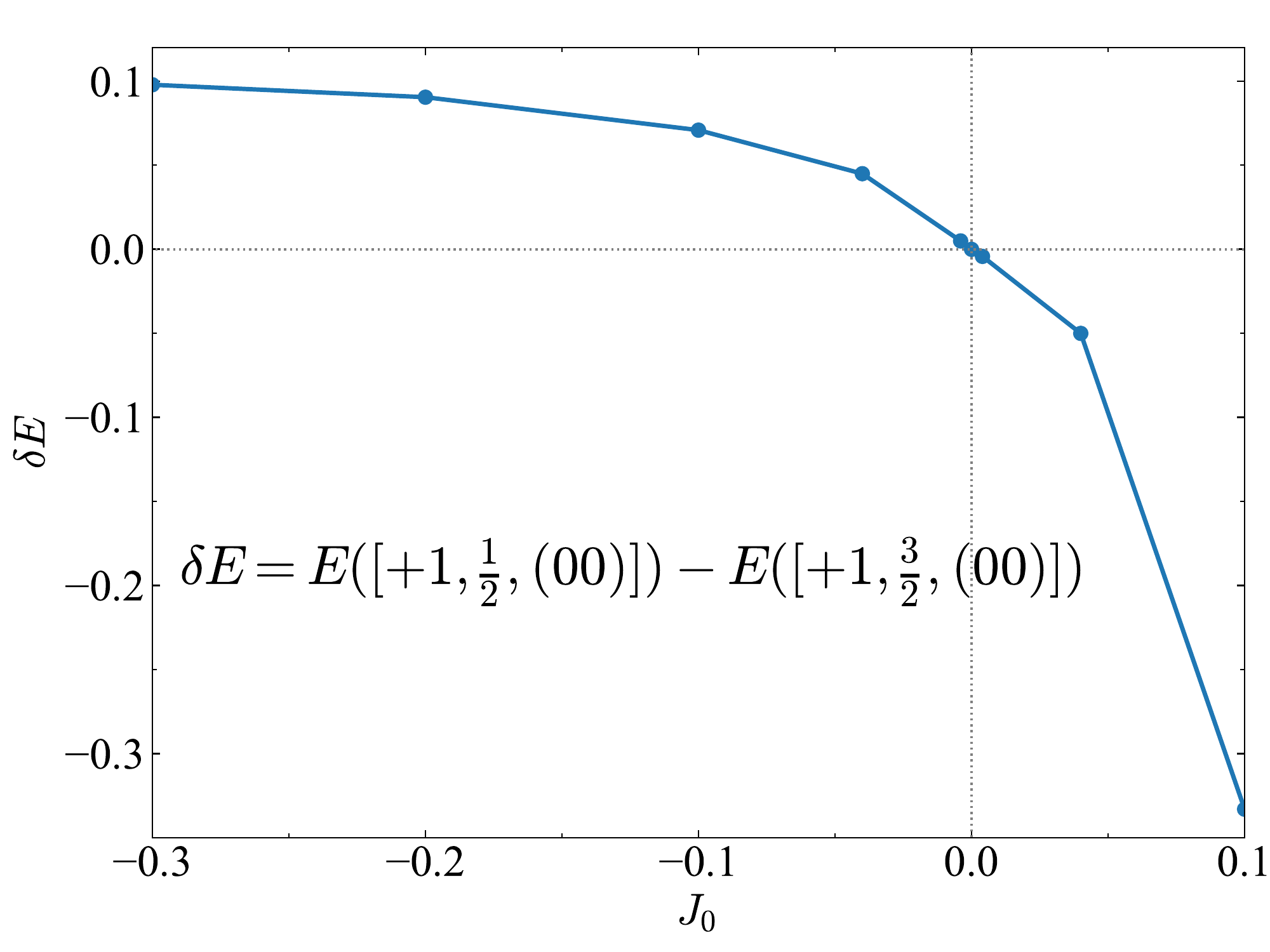}
    \caption{Energy difference of multiplets $[+1,\frac{1}{2},(00)]$ and $[+1,\frac{3}{2},(00)]$ along the phase boundary.}
    \label{fig:ediff}
\end{figure}

Fig.~\ref{fig:ediff} shows the energy difference $\delta E$
between the multiplets $[+1, \frac{1}{2}, (00)]$ and $[+1, \frac{3}{2}, (00)]$ as a function of $J_0$ along the phase boundary between FL and SFL (NFL). $I_0$ is fine-tuned at fixed $J_0$ from the FL side, to induce a very large crossover regime of $\it{SFL}^{\prime}$ or $\it{NFL}^{\prime}$.
The eigenlevel spectrum is taken at the odd Wilson site with the energy scale of $\omega_{k}=\Lambda^{-k/2}=10^{-8}$ and then the desired energy difference $\delta E$ is calculated. $[+1, \frac{3}{2}, (00)]$ has a lower energy when $J_0<0$, while $[+1, \frac{1}{2}, (00)]$ has a lower energy when $J_0>0$.

\bibliography{suppl}